\documentclass[a4paper,fleqn]{cas-dc}

\usepackage[numbers,sort&compress]{natbib} % Mathematical Biosciences uses numbered citations

\usepackage{amsmath}
\usepackage{amsthm}
\usepackage{nccmath}
\usepackage{mathtools}
\usepackage{bm}
\usepackage{graphicx}
\usepackage{float}
\usepackage{caption}
\usepackage{subfigure}

\usepackage[dvipsnames]{xcolor}
\usepackage{color}
\usepackage{colortbl}

\usepackage[normalem]{ulem}
\usepackage{enumitem}
\usepackage{blindtext}

\usepackage{url}
\usepackage{doi}

\usepackage[theorems,breakable]{tcolorbox}

\usepackage[capitalize,nameinlink]{cleveref}

\newcommand{\red}[1]{{\color{red}#1}}

\newcommand{\mathcolorbox}[2]{\colorbox{#1}{$\displaystyle #2$}}

\newtcbtheorem{mynote}{Process Flow}{%
    theorem name,%
    colback=blue!5,%
    colframe=blue!35!black,%
    fonttitle=\bfseries,
    title after break={Process Flow -- \raggedleft Continued}%
}{lem}

\restylefloat{table}

\usepackage{listings}
\definecolor{codegreen}{rgb}{0,0.6,0}
\definecolor{codegray}{rgb}{0.5,0.5,0.5}
\definecolor{codepurple}{rgb}{0.58,0,0.82}
\definecolor{backcolour}{rgb}{0.98,0.98,0.95}
\definecolor{backcolour2}{rgb}{0.99,0.99,0.98}

\lstdefinestyle{mystyle}{
    backgroundcolor=\color{backcolour},   
    commentstyle=\color{codepurple},
    keywordstyle=\color{blue},
    numberstyle=\tiny\color{codegray},
    stringstyle=\color{blue},
    basicstyle=\ttfamily\footnotesize\bfseries,
    breakatwhitespace=false,         
    breaklines=true,                 
    captionpos=t,                    
    keepspaces=true,                 
    numbers=left,                    
    numbersep=5pt,                  
    showspaces=false,                
    showstringspaces=false,
    showtabs=false,                  
    tabsize=2
}
\def\tsc#1{\csdef{#1}{\textsc{\lowercase{#1}}\xspace}}
\tsc{WGM}
\tsc{QE}
\tsc{EP}
\tsc{PMS}
\tsc{BEC}
\tsc{DE}

\usepackage{chngcntr}

\newtheorem{theorem}{Theorem}

\newtheorem{corollary}{Corollary} 
\newtheorem{lemma}{Lemma}

\theoremstyle{definition}
\newtheorem{definition}{Definition}

\newproof{pf}{Proof}
\newproof{pfs}{Proof Sketch}

\begin{document}
% Fix for cas-dc template version mismatch  
% Create \affiliation as alias to \address
\ExplSyntaxOn
\cs_set_eq:NN \affiliation \address
\ExplSyntaxOff

\let\WriteBookmarks\relax
\def\floatpagepagefraction{1}
\def\textpagefraction{.001}

\shorttitle{Forcing functions and identifiability}

\shortauthors{JR Conrad et~al.}

\title [mode = title]{Examining the impact of forcing function inputs on structural identifiability} 
\tnotemark[1]

\author[1,2]{Jessica Rose Conrad}[type=editor,
                        auid=000,bioid=1,
                        orcid=0000-0003-0008-0349]

\cormark[1]
\ead{jrconrad@umich.edu}

\affiliation[1]{University of Michigan, Department of Mathematics,
    530 Church St., 
    Ann Arbor,
    48109, 
    MI,
    USA}

\affiliation[2]{Los Alamos National Laboratory, T-6 Theoretical Biology Group, 1450 Central Ave., Los Alamos, 87544, NM, USA}

\author[2,4]{James M. Hyman}[type=editor,
                        auid=000,bioid=1,
                        orcid=0000-0001-5247-5794]
\cormark[4]
\ead{mhyman@tulane.edu}

\affiliation[4]{Tulane University, Department of Mathematics, 6823 St. Charles Ave., New Orleans, 70118, LA, USA}

\affiliation[3]{University of Michigan, Departments of Epidemiology and Complex Systems, 1415 Washington Heights, Ann Arbor, 48109, MI, USA}

\author[1,3]{Marisa C. Eisenberg}[type=editor,
                        auid=000,bioid=1,
                        orcid=0000-0003-4221-830X]

\cormark[1]
\ead{marisae@umich.edu}

\cortext[cor1]{Corresponding author}

% Abstract
\begin{abstract}
    Accurate parameter estimation in biological models---from optimizing drug dosing schedules to forecasting disease outbreaks—requires that model parameters be uniquely identifiable from available data.
    When parameters are not uniquely identifiable, estimation yields non-unique solutions, leading to unreliable predictions and misallocated resources.
    Existing approaches to resolving identifiability problems often involve re-parameterizing models, but this can reduce the biological interpretability of parameters. 

    We present a mathematical framework demonstrating that time-varying parameters, driven by environmental or behavioral data—such as temperature patterns, treatment schedules, or contact rates—can resolve parameter identifiability issues without re-parameterization.
    We prove that incorporating such forcing functions into model parameters cannot worsen structural identifiability, and establish conditions under which they provably improve it.
    Our results illustrate how forcing functions can be used to break parameter entanglements, with a single forcing function sometimes resolving multiple identifiability issues simultaneously.

    We validate this framework through pharmacokinetic models (drug tissue-blood exchange rates) and disease transmission models (seasonally varying contact rates), demonstrating that readily available time-varying data can resolve identifiability while preserving parameter interpretability.
    By connecting rigorous mathematical proofs to practical applications, this work provides modelers with systematic guidance on when, where, and how to leverage existing data streams to improve parameter estimation.
\end{abstract}


%\item Mathematical models with unidentifiable parameters produce non-unique estimates—forcing functions can resolve this without re-parameterizing, preserving the biological meaning of parameters
%\item We prove mathematically that incorporating time-varying forcing functions into parameters cannot degrade existing structural identifiability
%\item Additive parameter combinations separate when scaled; multiplicative and mixed structures resolve through strategic replacement
%\item One forcing function input can break multiple parameter entanglements simultaneously, enabling identification of entire connected parameter sets
%\item Our decision framework guides which parameters to target based on identifiable combination patterns, with examples from pharmacokinetics and disease modeling applications

% Keywords
\begin{keywords}
structural identifiability \sep 
practical identifiability \sep 
estimability \sep 
differential equations \sep 
forcing function \sep 
differential algebra \sep 
mathematical modeling \sep 
disease modeling \sep 
pharmacokinetics modeling \sep 
nonlinear model
\end{keywords}

\maketitle

\section{Introduction}

\noindent In biological systems, parameter estimation is a critical step in developing and using models for practical guidance for decision-making. When unique parameter estimates cannot be obtained for a model, there may be real-world consequences, such as misplaced supplies, poorly designed experiments, or unrealistic predictions.
\textit{Parameter identifiability} is a critical property to ensure successful parameter estimation and evaluates whether the optimal parameter estimates for a given model and data set or data type are unique.

Two broad (and sometimes overlapping) identifiability properties are typically considered: 1) structural identifiability, which assesses whether the model parameters are theoretically uniquely determinable given a particular measurement structure, without consideration of noise or bias, and 2) practical identifiability (sometimes also termed estimability), which assesses whether the model parameters are uniquely estimated given noisy, potentially incomplete data \cite{saccomani2003parameter, eisenbergDeterminingIdentifiableParameter2014, meshkat2015identifiability}.
Notably, recent approaches have emerged that integrate these concepts into a unified workflow, such as Profile-Wise Analysis (PWA) \cite{simpson2023profile}.

Both properties depend on the underlying model of the system and measurement process. Still, only practical identifiability accounts for issues such as noise, bias, and other problems arising from real-world, specific data sets.
Structural identifiability is a prerequisite for practical identifiability and is therefore necessary to ensure successful parameter estimation.

Identifiability problems have been observed in a range of biological and sociological models, including disease transmission models, pharmacokinetic models, population models of age-period-cohort effects, and engineering systems \cite{kao2018practical, roosa2019assessing,sefkow1996kinetic, cole2010determining, conrad2016modeling, bakshi1993optimal}. In fact, some have argued that unidentifiability and related concepts may be a generic or ubiquitous property across systems biology models \cite{gutenkunst2007sloppiness}.

Given the prevalence of identifiability issues in models across a range of fields, several approaches have been developed to improve identifiability, including: collecting more data \cite{raue2011addressing, kutalik2004optimal, balsa2008computational, raue2009structural}, re-parameterizing the model to reduce the number of parameters \cite{meshkat2009algorithm, meshkat2014finding, raue2011addressing}, restricting the parameter space with outside information \cite{raue2009structural}, or adding additional inputs if the system inputs can be changed \cite{villaverde2018input, meshkat2012alternative}.

Additional data collection can be expensive, and reparameterizing the model can alter the interpretability of its variables and parameters, sometimes in ways that hinder its intended use \cite{tikhomirov2008interiors, kao2018practical, chowell2017fitting, mordecai2013optimal}.

One approach that has received less attention from a formal analytical perspective is adding a time-varying driver or input function to the parameters (i.e., a parameter forcing function) to improve structural model identifiability.
While most explorations of the effect of inputs on identifiability have focused on the case where known inputs are added as a new term to the differential equation \cite{massonis2021structural, iliadis2019structural, distefano2015dynamic} or replacing a time-varying input with a constant parameter(s) \cite{villaverde2018input}, here we explore how inputs that are added as a time-varying parameter (either replacing or scaling an existing parameter) may impact identifiability. Recent theoretical work has also begun to extend observability rank conditions to time-varying nonlinear systems by using extended Lie derivatives to handle explicit time dependence \cite{martinelli2022extension}.

Such inputs are standard in cases where one has (or can make, in the case of engineered or controlled systems) a time-varying parameter of the model, such as a climate-driven transmission parameter for infectious disease models \cite{eisenberg2013examining, aldstadt2012space, chaves2012nonlinear, rotela2007space}, or a time-varying chemical reaction rate that depends on other substrates \cite{cepeda2019estimating, endalew2020flow, bakshi1993optimal}.
For instance, in epidemiological modeling, time-varying transmission rates are frequently used, and their impact on identifiability is a critical area of study \cite{gallo2022lack, cumsille2023general}.

Such time-varying parameters are often treated as constants for simplicity. Still, previous studies have shown that incorporating data on these time-varying parameters when available can improve both structural and practical identifiability.
For example, including parameter-forcing function inputs has been shown to improve the structural identifiability of disease transmission models with a seasonally forced contact rate \cite{evansStructuralIdentifiabilitySusceptible2005}, and to improve the practical identifiability of models of cholera transmission in Haiti when data on environmental transmission drivers, such as rainfall, are included \cite{eisenberg2013examining}.

\vspace{0.2cm}
\noindent In this study, we examine the impact of parameter forcing function inputs on structural identifiability. We show that incorporating this known/measured temporal variation into a parameter (either by replacing or scaling the parameter with the new input forcing function) will not worsen the structural identifiability of the model, and in some cases can substantially improve it. This work provides a theoretical framework specifically addressing how forcing functions affect structural identifiability.

This paper is structured as follows. \Cref{s:def} establishes the foundational framework, defining the model structure and key identifiability concepts. \Cref{s:diff_alg_method} then details the core analytical engine, the Differential Algebra Method, including Ritt's Pseudo-Division Algorithm, and discusses recent computational advancements. Building on this, \Cref{s:incorp} presents our first main theoretical result: that incorporating forcing functions cannot worsen structural identifiability. \Cref{s:forcefunc} follows with theorems demonstrating when these input functions improve identifiability. We then translate this theory into a practical guide in \Cref{s:method}, outlining a procedural approach for modelers. \Cref{s:generic_ex_main} provides a focused illustration using generic linear models. Finally, \Cref{s:examples} showcases the application of these concepts to real-world examples from the literature, before we conclude in \Cref{s:conclusions}.

\section{Framework and Definitions} \label{s:def}
\noindent This section lays the groundwork for our analysis of structural identifiability. We first define the general structure of ordinary differential equation (ODE) models (\Cref{s:model_structure}) used throughout the paper. We then introduce key concepts related to parameter identifiability, including global, local, and unidentifiability, as well as the notion of a forcing input function (\Cref{s:ident}). Finally, we discuss the utility of parameter graphs (\Cref{ss:paramgraph}) as a visual tool for understanding relationships between unidentifiable parameters.

\subsection{Model Structure} \label{s:model_structure}
\noindent We begin by introducing the model structure framework used in this paper. We will focus on ODE systems of the following general form:
\begin{equation} \label{eq:model}
    \begin{aligned}
    \dot{\pmb{x}} &=f(\pmb{x}(t, \pmb{\theta}),\pmb{u}(t),\pmb{\theta}),\\
    \pmb{x}(0, \pmb{\theta}) &= x_0(\pmb{\theta}), \\
    \pmb{y}(t, \pmb{\theta}) &= g(\pmb{x}(t, \pmb{\theta}), \pmb{\theta}),
    \end{aligned}
\end{equation}
where variable definitions can be found in \cref{tab:def}. 
We will refer to such a system as a model, denoted $M(\pmb{\theta})$.
A dot is placed above a function name to denote the time derivative of a function. Specifically, we will \emph{not} be using the tick symbol $'$ to indicate derivatives; instead, this will be used to distinguish different forms of equations, as described below.
$\pmb{u}(t)$ is a time-dependent vector of (known) input functions, if any.

All parameters $\pmb{\theta}$ and functions are assumed to be real-valued. 

Our goal is to introduce to $M(\pmb{\theta})$ a new known forcing input function $\tilde{u}(t)$. 
We will consider the effect of $\tilde{u}(t)$ where a parameter $\tilde{\theta} \in \pmb{\theta}$ in the model $M(\pmb{\theta})$ is either scaled as $\tilde{u}(t) \tilde{\theta}$ or replaced by $\tilde{u}(t)$.
We then evaluate if the newly generated model $\tilde{M}(\pmb{\theta})$ is structurally identifiable.

\begin{table}[!hptb]
    \centering
    \small
    \begin{tabular}{|c|p{6.2cm}|}
        \hline
         \textbf{Symbol} &  \textbf{Definition} \\
         \hline
         \multicolumn{2}{|c|}{\textit{Model Components}} \\
         \hline
         $\pmb{x}$  & State variable vector $\in \mathbb{R}^n$ \\
         \arrayrulecolor{lightgray}\hline
         $n$        & Number of state variables \\
         \arrayrulecolor{lightgray}\hline
         $t$        & Time \\
         \arrayrulecolor{black}\hline
         $\pmb{y}$  & Measured output vector $\in \mathbb{R}^r$ \\
         \arrayrulecolor{lightgray}\hline
         $r$        & Number of measured outputs \\
         \arrayrulecolor{black}\hline
         $\pmb{u}$  & Known input vector $\in \mathbb{R}^d$ \\
         \arrayrulecolor{lightgray}\hline
         $d$        & Number of inputs \\
         \arrayrulecolor{black}\hline
         $\pmb{\theta}$  & Parameter vector $\in \mathbb{R}^p$ (bold) \\
         \arrayrulecolor{lightgray}\hline
         $\theta_i$ & Individual parameter (non-bold) \\
         \arrayrulecolor{lightgray}\hline
         $p$        & Number of parameters \\
         \arrayrulecolor{black}\hline
         \multicolumn{2}{|c|}{\textit{Forcing Functions}} \\
         \hline
         $\tilde{u}$ & Forcing input being introduced \\
         \arrayrulecolor{lightgray}\hline
         $\tilde{\theta}$ & Parameter scaled/replaced \\
         \arrayrulecolor{black}\hline
         \multicolumn{2}{|c|}{\textit{Models \& Relations}} \\
         \hline
         $M(\pmb{\theta})$ & Model with parameters $\pmb{\theta}$ \\
         \arrayrulecolor{lightgray}\hline
         $\tilde{M}(\pmb{\theta})$ & Model after forcing function added \\
         \arrayrulecolor{lightgray}\hline
         $P(M)$ & Input-output relation from $M$ \\
         \arrayrulecolor{lightgray}\hline
         $P'(M)$ & Relation precursor (before monic) \\
         \arrayrulecolor{black}\hline
    \end{tabular}
    \caption{Notation conventions used throughout the manuscript.}
    \label{tab:def}
\end{table}

\subsection{Key Concepts} \label{s:ident}
\noindent To define identifiability, we first introduce the model map $\Phi: \pmb{\theta} \rightarrow \pmb{y}$, namely the map obtained by viewing the model as a map from the parameter space $\pmb{\theta}\in \mathbb{R}^p$ to the output space $\pmb{y}(t, \pmb{\theta}) \in \mathbb{R}^r$, given inputs $\pmb{u}(t)$ (where the map is determined by $f$, $x_0$, and $g$) \cite{eisenberg2013input, audoly2001global}.
The \textit{structural identifiability} of a model of the form of \cref{eq:model} is dependent on the injectivity of $\Phi$ \cite{eisenberg2013identifiability, meshkat2009algorithm, saccomani2003parameter, chowell2023structural}.

\begin{definition}[Globally Identifiable]
A model of the form of \cref{eq:model} is \textit{globally identifiable} if for almost every value of $\pmb{\theta}$ and almost all initial conditions, the equation $\pmb{y}(t, \pmb{\theta}) = \pmb{y}(t, \pmb{\overline{\theta}})$ implies $\pmb{\theta} = \pmb{\overline{\theta}}$ \cite{eisenberg2013identifiability}.
\end{definition}

If our system is globally identifiable, then the model output $\pmb{y}(t,\pmb{\theta})$ uniquely determines $\pmb{\theta}$.

\begin{definition}[Locally Identifiable]
If $\pmb{y}(t, \pmb{\theta}) = \pmb{y}(t, \pmb{\overline{\theta}})$ implies $\pmb{\theta} = \pmb{\overline{\theta}}$ within an $\epsilon$-neighborhood of $\pmb{\theta}$, 
then the model is \textit{locally identifiable} \cite{evansStructuralIdentifiabilitySusceptible2005}. Note that global identifiability implies local identifiability.
\end{definition}

\begin{definition}[Unidentifiable]
The model is said to be \textit{unidentifiable} if it is not locally identifiable.
\end{definition}

If a model is unidentifiable,  a subset of parameters $\tilde{\pmb{\theta}} \subset \pmb{\theta}$ in the model may still be identifiable. 
The remaining subset of unidentifiable parameters $\pmb{\theta}\setminus \{\tilde{\pmb{\theta}} \}$ may instead contain \textit{identifiable parameter combinations}. 
These are the functional forms of dependencies between unidentifiable parameters \cite{eisenbergDeterminingIdentifiableParameter2014, bhola2023estimating}. 
If identifiable parameter combinations are known, the model can be re-parameterized to be at least locally identifiable.

\begin{definition}[Forcing Input Function]
A \textit{forcing input function} is a bounded, non-constant, $C^n$-smooth function (where $n$ is the dimension of the state space), whose value is known exactly for all time values in the interval of interest. The $C^n$ requirement ensures sufficient differentiability for the calculation of the characteristic set through Ritt's pseudo-division algorithm.
\end{definition}

In practice, forcing functions are often assumed to be $C^\infty$ (infinitely differentiable) or analytic to avoid concerns about differentiability order.

We also assume that inputs are consistent with the solvability assumption used for input-output equations, as discussed further below.

\setlength{\leftskip}{0.5cm}
\vspace{0.2cm}
\noindent \textit{Remark: }
    This definition of a forcing input function does not include commonly used classes of inputs such as Dirac delta functions (often interpreted as initial conditions) or constants \cite{audoly2001global, eisenbergDeterminingIdentifiableParameter2014, meshkat2015identifiability}. 
    These are often insufficiently time-varying for our purposes.
    Here, references to a `forcing input function' refer to the case where a sustained, non-constant function drives the system throughout the time interval of interest.
\setlength{\leftskip}{0cm}
\medskip

\noindent \textbf{Assumption 1.}
Throughout this work, we assume all parameters of interest appear in at least one input-output equation derived from the model. Parameters that do not appear in any input-output equation (that is, parameters that do not affect the measured outputs) cannot be identified from output measurements alone, regardless of whether forcing functions are incorporated. Such parameters exhibit structural non-identifiability at a fundamental level and require either additional measurements or a reformulation of the model or considerations of initial conditions.
%}

\subsection{Parameter Graph} \label{ss:paramgraph}
\noindent To motivate the discussion of when to choose gathering data on a parameter(s) over re-parameterizing the model, we review parameter graph methodology. 
Parameter graphs visualize the relationships among parameters, providing guidance on which parameters can be targeted to improve identifiability and estimability.
Parameters are represented as nodes in a hypergraph, with identifiable combinations serving as the edges \cite{eisenbergDeterminingIdentifiableParameter2014}.

\begin{figure}
    \centering
        \includegraphics[width=0.2\textwidth]{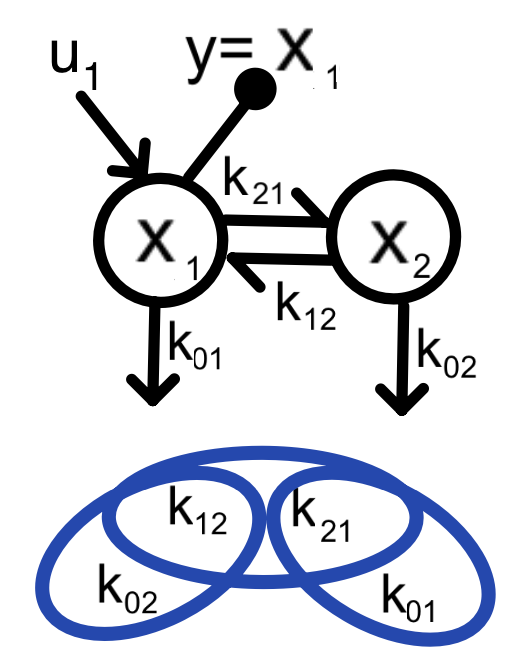}
        \captionof{figure}{An example visualization of parameter combinations (bottom) for a 2-compartment model (top). Circles indicate the three identifiable parameter combinations, e.g. $k_{02}$ and $k_{12}$ are involved in an identifiable combination. Some parameters, such as $k_{12}$, are involved in more than one combination. In this case, one connected component has been identified for the system [\citenum{eisenbergDeterminingIdentifiableParameter2014}].}
    \label{fig:paramgraph}
\end{figure}

Even without the exact mathematical equations for the 2-compartment model shown in \cref{fig:paramgraph} (presented later in \Cref{s:diff_alg_method}), we can see directly that all the parameters in such a model might be connected through identifiable parameter combinations.
If even a single parameter $k_*$ in such a connected component were measured or known, all other parameters within that component could potentially become unique.

That is, we can visually demonstrate that for each connected component in a parameter graph of size > 1, at least one parameter must be measured or known to make all other parameters in the component unique.
This can be used as a guide for when and where to apply the theorems outlined in this paper.

Applying measured data on targeted parameters can potentially improve the structural and practical identifiability of a model without re-parameterizing.
While parameter graphs can help identify which parameters to target, they do not reveal the analytical form of the parameter combination. To reveal the analytical form, we perform further analysis by solving the input-output equation(s) to derive the coefficient map, a process detailed in \Cref{s:diff_alg_method}. 

In the following sections, we provide a mathematical roadmap for identifying specific parameters within functional parameter combinations that can be targeted for scaling and replacement. 
A practical summary of this roadmap is given in \Cref{s:method}.

\section{Differential Algebra Method} \label{s:diff_alg_method}
\noindent This section details the core mathematical methodology used in this paper to analyze structural identifiability. We begin by illustrating the goal of this approach with a brief example that transforms a system of ordinary differential equations into an input-output relation. We then formally describe the characteristic set algorithm (\Cref{RPDalgorithm}), a key technique from differential algebra for achieving this transformation. Finally, we present a motivating pharmacokinetics example (\Cref{ss:MotEx2_diff_alg}) to further illustrate these concepts and their relevance to identifiability analysis in a practical biological context.

Before detailing the algorithmic steps, let's consider a simple illustrative example to motivate the process. Consider a 2-compartment model described by:
\begin{equation}\label{eq:smallexample}
\begin{aligned}
     \dot{x}_1 & = -k_{21} x_1 \\
     \dot{x}_2 & = k_{21} x_1 - k_{02} x_2 \\
     y & = x_2
\end{aligned}
\end{equation}
where $x_1, x_2$ are unobserved states, $y$ is the measured output, and $\pmb{\theta} = \{k_{21}, k_{02}\}$ are parameters.
The goal of the differential algebra approach is to systematically eliminate the state variables to derive an \textit{input-output relation} solely in terms of $y$ and its derivatives, and $\pmb{\theta}$. For this system, the derived input-output relation is:
$$ \ddot{y} + (k_{02} + k_{21})\dot{y} + k_{02}k_{21} y = 0 $$
This equation, and its coefficients, are then used to assess the structural identifiability of $k_{21}$ and $k_{02}$. As shown in \Cref{RPDexample}, this original system is locally identifiable, but not globally identifiable. 

While we could impose a constraint on the system to resolve this identifiability issue, alternatively, we can introduce new information in the form of a forcing function.
If a parameter like $k_{21}$ is scaled by a known forcing function $u(t)$, the method yields the more complex input-output relation (terms ordered to emphasize the comparison):
%$$ - df(u,t)*df(y,t) - df(u,t)*y*k02 + df(y,t,2)*u + df(y,t)*u**2*k21 + df(y,t)*u*k02 + u**2*y*k02*k21$$
$$\mathcolorbox{Yellow}{u} \ddot{y}+ (k_{02}\mathcolorbox{Yellow}{u} + k_{21} \mathcolorbox{Yellow}{u^2} - \mathcolorbox{Yellow}{\dot{u}})\dot{y} 
+ (k_{02}k_{21}\mathcolorbox{Yellow}{u^2} - k_{02}\mathcolorbox{Yellow}{\dot{u}}) y = 0$$
This new system is globally identifiable, and we can see that incorporating $u(t)$ changes the calculation of the input-output equation such that the analogous terms in the previous equation are now multiplied by $u$ and in some cases split (e.g. what was $(k_{02}+k_{21})\dot{y}$ has now become two distinct monomials, $k_{02}u\dot{y}$ and $k_{21}u^2\dot{y}$). Additional terms are also added due to the derivatives taken of the $u$ variable (e.g. $\dot{u}y$). The detailed step-by-step application of the characteristic set algorithm to both the original and augmented versions of this example is provided in \Cref{RPDexample}. We now proceed to describe the general algorithm.

The differential algebra method can be implemented automatically for any rational function differential equation model through the software DAISY (Differential Algebra for Identifiability of Systems) \cite{bellu2007daisy}. However, for complex models, this method can become computationally intractable \cite{saccomani2003parameter}. Recent advancements in numerical algebraic geometry offer solutions for larger and more complex systems where symbolic methods struggle \cite{bates2019identifiability}.
More recent and computationally efficient tools based on differential elimination and Gr\"{o}bner bases, such as SIAN \cite{hong2019sian} and StructuralIdentifiability \cite{dong2023differential}, also perform this task by computing the characteristic set or input-output equations without requiring manual steps. For exact details on this algorithm, please refer to \citet{ritt1950differential}, \citet{bellu2007daisy}, and \citet{saccomani2003parameter}. 

The differential algebra method can be used to determine not only the model's identifiability but also the identifiable parameter combinations \cite{eisenberg2013input}.
The overall goal is to reduce the system of differential equations to an equation involving only known or measured variables and parameters, referred to as the \textit{input-output relation}.

\begin{definition}[Input-Output Relation]
\textcolor{black}{An input-output relation is a monic polynomial differential equation of the form $\Psi(y, u, \pmb{\theta}) = 0$, where $\Psi$ represents the characteristic set polynomial(s) obtained by eliminating all state variables from the original ODE system $M$. An input-output relation involves only the measured outputs $y$, known inputs $u$, their time derivatives, and the unknown parameters $\pmb{\theta}$. Typically, for a given model $M$, the number of input-output relations is equal to the number of measured output variables.}
\end{definition}

\noindent\textbf{Notation}. We will also define some new terminology around input-output relations.
For any model $M$ of the form defined above, let $P(M)$ be an input-output relation derived from $M$, and let $P'(M)$ be the precursor to $P(M)$ in the characteristic set algorithm such that $P'(M)$ becomes $P(M)$ when divided by the coefficient of its leading monomial.

\vspace{0.2cm}
\noindent The input-output relation is simply a higher-order representation of the original first-order model $M(\pmb{\theta})$, now without the state variables.
The coefficients of the input-output relations can be used to test the model's identifiability and to determine identifiable parameter combinations as necessary \cite{eisenberg2013input}.

\begin{figure}
    \centering
    \includegraphics[width=0.25\textwidth]{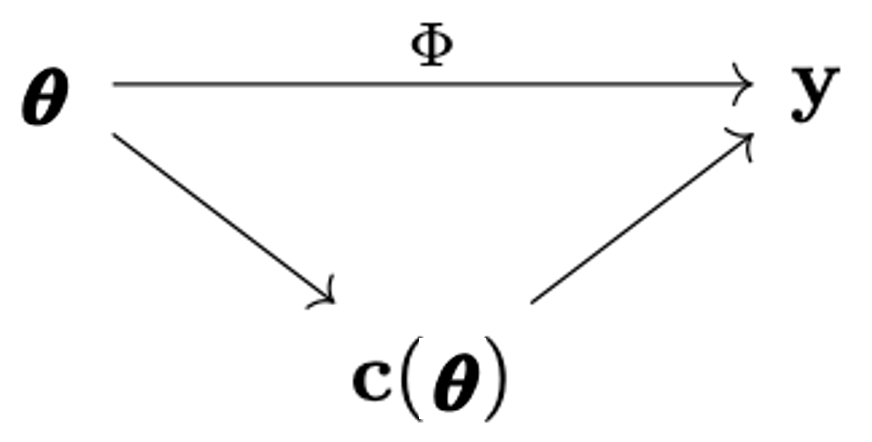}
    \caption{Map of relationships between the parameter space $\pmb{\theta}$ and the measured output(s) $y$. Analysis reveals that both $\Phi$ and $c(\pmb{\theta})$ represent maps from the parameter space to the output space, indicating either can be used to determine the structural identifiability of $M(\pmb{\theta})$.} 
\label{fig:commutemap}
\end{figure}

\setlength{\leftskip}{0.5cm}
\vspace{0.2cm}
\noindent \textit{Remark. }
    If we take $c(\pmb{\theta})$ to be the coefficients of the input-output relation, we can write the following commutative diagram shown in \cref{fig:commutemap}.
    The map $c(\pmb{\theta}) \rightarrow y$ is usually assumed to be injective, an assumption termed \textit{solvability}, defined further below.
    Therefore, to evaluate the injectivity of the model map $\Phi$, we can instead evaluate the coefficient map $\pmb{\theta} \mapsto c(\pmb{\theta})$ \cite{eisenberg2013input}.
    
\setlength{\leftskip}{0cm}

\begin{definition}[Parametric Solution Set]
The fibers of the coefficient map $c(\pmb{\theta})$ (or equivalently of the model map $\Phi$) are termed the parametric solution sets of the system. These are all parameter values that generate a particular output trajectory or a set of coefficient values.
\end{definition}

\begin{definition}[Solvability]
\textit{Solvability} is the assumption that we have sufficiently many \textit{independent} values for $\pmb{y}$ and $\pmb{u}$ and their derivatives so that these known values enable us to form a linear system from the input output relation which is solvable for the the coefficients (the unknown variables in this case) \cite{saccomani2003parameter}. 
\end{definition}

This is typically the case for most structural identifiability purposes, as we are assumed to have noise-free, continuous measurements of $\pmb{y}$ and $\pmb{u}$ (so we have arbitrarily many points, which generically should be independent). However, there are situations in which the solvability assumption is violated, typically when there is a constraint on the variables, inputs, or outputs that implies that $\pmb{y}$ and $\pmb{u}$ lie in a lower-dimensional subspace. 
In such cases, the constraint should typically be incorporated into the model (often simplifying it) so that the solvability assumption is met.
\medskip

\noindent \textbf{Assumption 2.}
We assume that the introduced forcing functions $\tilde{u}(t)$ satisfy the solvability assumption. Specifically, $\tilde{u}$ and its derivatives must not introduce additional constraints that reduce the effective rank of the measurement system. For example, if $\tilde{u}(t)$ were constrained to lie on a lower-dimensional manifold related to the outputs $\pmb{y}$, this could violate solvability. In practice, forcing functions derived from independent physical processes (e.g., climate data, experimental inputs) typically satisfy this requirement.
%}

\subsection{Characteristic set calculation using Ritt's pseudo-division} \label{RPDalgorithm}

\noindent To calculate the characteristic set using Ritt's pseudo-division, model variables (inputs, outputs, states, and their derivatives) must be ranked. 
\citet{bellu2007daisy} use a standard ranking system that declares input and output variables as the lowest ranked, with the derivatives further ranked as follows: 
\begin{multline} \label{ranking}
u_1 < u_2 < ... <\dot{u}_1 < \dot{u}_2 < ... \\< y_1 < y_2 < ... <\dot{y}_1 < \dot{y}_2 < ...\\< x_1 < x_2 < ...<\dot{x}_1 < \dot{x}_2 < ...
\end{multline}

Note that any ranking that is consistent with the calculation of the characteristic set will work, provided that forcing inputs are ranked below outputs and state variables. 
%Alternative rankings with $\tilde{u}$ lowest-ranked (below outputs and states) also preserve our results, but we use this standard ranking throughout for clarity and consistency.

Let $w_j$ be the \textit{leader} of the polynomial $A_j$ such that it is the highest ranking derivative of the variables appearing in that polynomial \cite{meshkat2012alternative, bellu2007daisy}.
A polynomial $A_i$ is \textit{lower rank} if $w_i < w_j$, or if $w_i = w_j$ then the algebraic degree of $A_i$ is less than that of $A_j$. 
$A_i$ is \textit{reduced with respect to polynomial $A_j$} if $A_i$ does not contain the leader of $A_j$ with equal or greater algebraic degree, nor its derivatives.

\setlength{\leftskip}{0.5cm}
\vspace{0.2cm}
\noindent \textit{Remark.} 
The equations of our system $M$ need to be transformed into polynomials $A_{i \in \mathbb{N}}$ that are equivalent before we can proceed. 
This will allow us to compare polynomials as described above. 
Therefore, before proceeding with the algorithm, we will impose the following step: 
\begin{enumerate}[start=0]
    \item \textbf{Rewrite:} Set all equations of the model $M(\pmb{\theta})$ equal to zero such that all nonzero terms are gathered to the right of any equal sign. 
    Order the polynomials by increasing rank, such that the highest-ranked polynomial is $A_{r+n}$. 
\end{enumerate}
\setlength{\leftskip}{0cm}

We can now iterate through the characteristic set algorithm using Ritt's Pseudo-Division, ensuring that all polynomials in the system are reduced with respect to one another.
We traverse the following loop:
\begin{enumerate}
    \item \textbf{Compare:} Define $A_1$ as our first $A_j$ polynomial. If $A_j$ is fully reduced compared to all other polynomials in the set, choose the polynomial one rank higher than the current $A_j$, and make it the new $A_j$.

    Compare the leader of $A_j$ to other higher-ranked polynomials to select $A_i$.
    If the leader of a higher-ranked polynomial in the set is the $k$-th derivative $w_j^{(k)}$ of the leader of $A_j$, let this be $A_i$. 
    
    Differentiate $A_j$ $k$-times to find $A_j^{(k)}$ with leader $w_j^{(k)}$.
    
    \item \textbf{Reduce:} Multiply the polynomial $A_i$ by the coefficient of the highest power of $w_j^{(k)}$ and let $R$ be the remainder of the division of this new polynomial by $A_j^{(k)}$ with respect to the variable $w_j^{(k)}$. Then $R$ is reduced with respect to $A_j^{(k)}$. The polynomial $R$ is called the pseudo-remainder of the pseudo-division.
    \item \textbf{Replace:} The polynomial $A_i$ is replaced by the pseudo-remainder $R$ and the sub-loop (Step 2 and 3) is iterated using $A_j^{(k-1)}$ in place of $A_j^{(k)}$ and so on, until the pseudo-remainder is reduced with respect to $A_j$.
    \begin{enumerate}
        \item \textit{Repeat Sub-loop:} Check that $A_j$ is now reduced against all higher ranked polynomials. If a higher-ranked polynomial is not fully reduced with respect to $A_j$, label this polynomial $A_i$ and repeat Steps 2 and 3.
        \item \textit{Exit Sub-loop:} When all higher-ranked polynomial(s) $A_i$ are reduced with respect to $A_j$, reorder the polynomials by increasing rank and return to Step 1.
    \end{enumerate}
\end{enumerate}
The result is a family of differential polynomials belonging to the differential ring $R(\pmb{\theta})[\pmb{u},\pmb{y},\pmb{x}]$ as shown in \cref{ring}.
\begin{align} \label{ring}
    A_1(\mathbf{u},\mathbf{y}), ..., A_r(\mathbf{u},\mathbf{y}) \\ \nonumber
    A_{r+1}(\mathbf{u},\mathbf{y},\mathbf{x_1}), \\ \nonumber
    A_{r+2}(\mathbf{u},\mathbf{y},\mathbf{x_1}, \mathbf{x_2}), \\ \nonumber
    \vdots ~~~~~~~~~\\ \nonumber
    A_{r+n}(\mathbf{u},\mathbf{y},\mathbf{x_1}, ..., \mathbf{x_n})
\end{align}
The first $r$ differential polynomials $A_i$, $i = 1:r$ of \cref{ring} do not depend on $\pmb{x}$, and  form the precursor to the \textit{input-output relation} of the system, $P'(M)$ \cite{bellu2007daisy}.
It always takes $n$ iterations of the Ritt's Pseudo-Division loop (Steps 1 to 3) to produce the precursor input-output relation for the system.

To ensure that the input-output relation polynomials are unique at the end of this process, we enforce the following final step \cite{bellu2007daisy}:
\begin{enumerate}[start=4]
    \item \textbf{Enforce Monic $\pmb{A_i, i = 1:r}$:} Check that the new polynomials $A_i, i = 1:r$ are monic. If not, divide each $A_i$ by its leading coefficient to enforce that $A_i$ are monic polynomials.
\end{enumerate}
This generates our input-output equations, $P(M)$, where the coefficients represent the identifiable combinations of $M$.

\vspace{0.2cm}
For a detailed example of how to apply Ritt's Pseudo-Division to an ODE system of interest, please refer to \cref{RPDexample}.

\subsection{Motivating Example: Pharmacokinetics} \label{ss:MotEx2_diff_alg}

\noindent Consider the following system of equations from \citet{eisenbergDeterminingIdentifiableParameter2014} that is illustrated in more detail in \cref{fig:paramgraph} (from \Cref{s:def}, now with a slightly modified $y$ equation):
 \begin{align} 
     \dot{x}_1 & = k_{12} x_2 - (k_{21} + k_{01}) x_1 + u_1  \label{ex2_diff_alg} \\  \nonumber
     \dot{x}_2 & = k_{21} x_1 - (k_{12} + k_{02}) x_2; \\ \nonumber
     y & = x_1 / V 
 \end{align}
This is a standard model used in pharmacokinetics, where $x_1$ is the mass of the drug in the blood (e.g., drug dose), and $x_2$ is the subsequent mass of the drug in the tissue.
The drug can exchange between blood and tissue ($k_{21}$ and $k_{12}$), and is degraded in both compartments ($k_{01}$ and $k_{02}$).
The measured output, $y$, is the drug concentration in blood, given the blood volume $V$ \cite{distefano2015dynamic,eisenbergDeterminingIdentifiableParameter2014, audoly2001global, meshkat2009algorithm}.

In this case, $u_1$ denotes the drug input function, which may be time-dependent.
This model has already been shown in the literature to be unidentifiable, but we briefly review the input-output relation for this example and expand on possible ways to make this system identifiable.

The exact derivation of the input-output relation for this system can be found in \citet{meshkat2009algorithm} or \citet{audoly2001global}.
The input-output relation for \cref{ex2_diff_alg} is:
\begin{multline} \label{pharmio_diff_alg}
    0 = \ddot{y} + (\mathcolorbox{LightYellow}{k_{01} + k_{21}} + \mathcolorbox{PowderBlue}{k_{02} + k_{12}})\dot{y} \\ ~~~~~~~~~ + ((\mathcolorbox{LightYellow}{k_{01} + k_{21}})(\mathcolorbox{PowderBlue}{k_{02} + k_{12}})  - k_{21}k_{12}) y \\- {(\mathcolorbox{PowderBlue}{k_{02} + k_{12}})} u_1 /{\mathcolorbox{Lavender}{V}}   - \dot{u}_1/{\mathcolorbox{Lavender}{V}} 
\end{multline}

%\textcolor{green}{From Daisy output 11/14: 
%$$- df(u,t) + df(y,t,2)*v + $$
%\\
%$$df(y,t)*v*(k01 + k02 + k12 + k21) - u*(k02 + k12) $$
%\\
%$$+ y*v*(k01*k02 + k01*k12 + k02*k21)$$}

Then the coefficient map is $c(\pmb{\theta}) = \{(k_{01} + k_{21} + k_{02} + k_{12}),  (k_{12}k_{21} - (k_{02} + k_{12})(k_{01} + k_{21})), (k_{02} + k_{12})/V, 1/V\}$.
Observe the recurring highlighted functional groups in \cref{pharmio_diff_alg}.
The coefficient map can be further reduced to show that identifiable parameter combinations are summarized by this set \cite{distefano2015dynamic}: $$c(\pmb{\theta}) = \{V, k_{12}k_{21}, k_{02} + k_{12}, k_{01} + k_{21} \}$$
While $V$ is obviously identifiable, the other parameters remain entangled in different parameter combinations.
As such, the system is not identifiable.

To make this model identifiable, it has been recommended to re-parameterize this model using the identifiable parameter combinations, setting $\{q_1, q_2, q_3, q_4\} = \{V,$ $k_{12}k_{21},$
$k_{02} + k_{12}, k_{01} + k_{21} \}$, and $x'_2 = k_{12}x_2$ \cite{meshkat2009algorithm}.
This yields the following revised system:
 \begin{align} 
     \dot{x}_1 & = -q_4 x_1 + x'_2 + u_1 \label{ex2b_diff_alg} \\ \nonumber
     \dot{x'}_2 & = q_2 x_1 - q_3 x'_2; \\ \nonumber
     y & = x_1 / q_1 
 \end{align}
While analysis reveals that this new system is globally identifiable, how can the new parameters be interpreted?
The parameter $q_2$ is now the \textit{product} of the flows between the blood and tissue, while $q_3$ and $q_4$ are the total rate of flows out of the blood and tissue, respectively.

Suppose modelers were interested in the question of how fast the drug is decaying in the tissue (or in general, the exact values and levels of the drug in the tissue). In that case, this question cannot be answered by the re-parameterized model in \cref{ex2b_diff_alg}, because the rate is now indistinguishable from the exchange rate of substances moving from the blood into tissue.
Mathematically, \cref{ex2b_diff_alg} represents an improved model, but for practical applications, \cref{ex2_diff_alg} remains, for the moment, the better choice for researchers.
We will return to the question of how to `improve' \cref{ex2_diff_alg} while maintaining interpretability in \Cref{ss:revisit}.

\vspace{0.2cm}

\noindent Having established the mathematical framework and methodology, we now turn to our main theoretical contributions. The following section presents our first key result: that forcing functions provide a safe approach to potentially improving identifiability.

\section{Incorporating forcing functions cannot worsen structural identifiability} \label{s:incorp}

\noindent This section rigorously establishes our first key result: that introducing forcing functions does not compromise a model's existing structural identifiability. Our argument unfolds in stages. We begin by proving two lemmas, which address how parameter solution sets behave when a forcing function is introduced, specifically showing that no new, problematic solutions are generated. With this foundation, Theorem~\ref{nothinglostthm} then formally states that identifiable parameters remain identifiable, and its corollaries then extend this finding to the identifiability of the overall model. Importantly, we only consider structural rather than practical identifiability here (since, for example, a highly noisy forcing function could potentially hinder practical estimation of otherwise structurally identifiable parameters, as discussed in \cite{murphy2024measurement}).

To fully understand the impact on identifiability, we must consider how the derivation of the input-output equation is affected. Lemma~\ref{newform} addresses this next by showing that the fundamental steps of the characteristic set algorithm using Ritt's Pseudo-Division remain consistent. Coefficients in the input-output equation precursor $P'(M)$ are preserved in the augmented $P'(\tilde{M})$, in a modified form. Other terms may be added as well, however, the original terms will remain under the modifications described below.

\textcolor{black}{A sketch of the proof for Lemma~\ref{newform} is shown here. For a full detailed proof, refer to \Cref{prop1proof}.}

\begin{lemma} \label{newform} 
Let $\tilde{u}$ be a forcing input function, and let $M$ be any rational function model of the form of \cref{eq:model}, assuming the standard ranking described in \cref{ranking}. Generate $\tilde{M}$ by scaling any parameter $\tilde{\theta} \in \pmb{\theta}$ in the state variable equations of $M$ with $\tilde{u}$.
     
Then for each input-output relation precursor equation $P'(M)$, $\tilde{M}$ will have an analogous input-output equation precursor $P'(\tilde{M})$, where the terms of $P'(M)$ will also be present in $P'(\tilde{M})$ with two modifications: 1) the replacement rule $\tilde{\theta} \mapsto \tilde{\theta}\tilde{u}$ (which may split a monomial into two or more distinct terms), and 2) the multiplication of all such analogous terms in $P'(\tilde{M})$ by a common term in the parameters and $\tilde{u}$. We describe these modifications in more detail below.

Let $C_im_i = (c_0 + c_1\tilde{\theta} + \cdots c_n\tilde{\theta}^n)m_i$ be a monomial term in $P'(M)$, where $C_i$ is a coefficient polynomial in the parameters (potentially including $\tilde{\theta}$, written here as a polynomial in $\tilde{\theta}$ where some $c_i$ may be zero), and  $m_i$ is a monomial in outputs, inputs, and their derivatives. 

Then $P'(\tilde{M})$ will contain analogous terms to $C_im_i$, of the form $ac_0m_i$, $ac_1\tilde{\theta}\tilde{u}m_i$, $\dots$ $ac_n\tilde{\theta}^n\tilde{u}^nm_i$, where $a$ is a common multiplicative term given by a polynomial in the parameters multiplied by a monomial in $\tilde{u}$ ($a$ may be trivial, e.g. $a=1$).

Additional terms not descended from any original term may also be generated in $P'(\tilde{M})$, but these will never be the leading term of $P'(\tilde{M})$.

\end{lemma}

\noindent
\noindent \textit{Remark.} 
It is important to note that the analogous terms in $P'(\tilde{M})$ will typically not all have identical monomials in $\tilde{u}$. For example, consider how the original coefficient $(k_{02} + k_{21})\dot{y}$ transforms when $k_{21}$ is scaled by $u$ in the simple example in Eq.~\eqref{eq:smallexample}. After applying the characteristic set algorithm with the scaling $k_{21} \mapsto k_{21}u$, this single term splits into multiple analogous terms: $k_{02}u\dot{y}$ (with $u^1$) and $k_{21}u^2\dot{y}$ (with $u^2$). While these have different powers of $u$, both:
\begin{itemize}
\item Form analogous terms to the same coefficient in $P'(M)$
\item Follow the replacement rule $k_{21} \mapsto k_{21}u$
\item Are then also multiplied by the same accumulated  coefficient $u$.
\end{itemize}

This splitting of terms with different $\tilde{u}$ powers is precisely what enables the separation of previously entangled parameters into distinct coefficient equations, as demonstrated in the results below. For a detailed worked example demonstrating this mechanism, see the 2-compartment model derivation in Appendix \ref{asec:2compex}.

\begin{pfs} 
    Let $M$ be any model of the form of \cref{eq:model}. 
    We take the following ranking on the input, output, and state variables as described previously:
    \begin{multline*}
        u_1 < u_2 < ... < \dot{u}_1 < \dot{u}_2 < ...\\  < y_1 < y_2< ... <\dot{y}_1 < \dot{y}_2 < ...\\  < x_1 < x_2 < ...<\dot{x}_1 < \dot{x}_2 < ...
    \end{multline*}
    Assume $\tilde{u}$ is any known bounded $n$-differentiable time-dependent forcing input function.
    Scale any parameter $\tilde{\theta} \in \pmb{\theta}$ in the state equations of $M$ by $\tilde{u}$ to create a new model $\tilde{M}$.
    For our ranking of the input, output, and state variables for the new system, we will choose the following:
    \begin{multline*}
        \mathcolorbox{Yellow}{\tilde{u}} < u_1 < u_2 < ...< \mathcolorbox{Yellow}{\dot{\tilde{u}}} < \dot{u}_1 < \dot{u}_2 < ... < ...\\  < y_1 < y_2 < ... <\dot{y}_1 < \dot{y}_2 < ... \\< x_1 < x_2 < ...<\dot{x}_1 < \dot{x}_2 < ...
    \end{multline*}
    where the new forcing input function is ranked lowest among the input variables, and importantly, less than output variables and state variables (noting that any ranking chosen consistent with calculating the characteristic set will also work so long as $\tilde{u}$ is ranked lowest). 
    
    We review the changes this introduction makes to each step of the characteristic set calculation: Rewrite, Compare, Reduce and Replace, and Enforce Monic $A_i$, by induction on the number of algorithm iterations.
    
    \textbf{Base Case (k=0):} This is given by the Rewrite step, where we observe that scaling a parameter $\tilde{\theta} \mapsto \tilde{\theta}\tilde{u}$ only affects coefficients in the polynomial form of the equations. Consider a term $\tilde{\theta} \cdot g(\pmb{x})$ in $M$; after scaling in $\tilde{M}$ this becomes $\tilde{\theta}\tilde{u} \cdot g(\pmb{x})$. The structure is preserved with the modification $\tilde{\theta} \mapsto \tilde{\theta}\tilde{u}$.
    
    \textbf{Inductive Step:} Assume the property holds for iteration $k$ of the characteristic set algorithm. We examine what happens when taking derivatives in the $(k+1)$-th step using the Leibniz rule.
    
    %For a term $\tilde{\theta}\tilde{u} \cdot h(\pmb{x})$ where $h$ depends on state variables, applying the Leibniz rule:
    %$$\frac{d}{dt}[\tilde{\theta}\tilde{u} \cdot h] = \tilde{\theta}\dot{\tilde{u}} \cdot h + \tilde{\theta}\tilde{u} \cdot \dot{h}$$
   
    When taking the $m$-th derivative using Leibniz rule, the term corresponding to $m$ in the binomial expansion preserves the original structure (with $\tilde{\theta} \mapsto \tilde{\theta}\tilde{u}$ modification, forming the analogous term(s) to $M$), while additional terms involving derivatives of $\tilde{u}$ are generated but ranked lower. This maintains the reduction order and ensures the modified original monomials are generated at the next step.
    
    In this manner, the modified original terms and their derivatives from $M$ are represented in $\tilde{M}$ during the steps of characteristic set calculation using Ritt's Pseudo-Division.
    Other terms involving derivatives of $\tilde{u}$ may be generated but will not alter the choice of leader for reduction steps and its derivatives \textcolor{black}{as these new terms will always be lower ranked. $\qed$}    
\end{pfs}

\vspace{0.2cm}
\noindent Lemma~\ref{newform} establishes that the characteristic set algorithm preserves structural relationships when introducing forcing functions. We now show that this preservation translates directly to containment of parameter solution sets, which is the foundation for proving that forcing functions cannot worsen identifiability.

\begin{lemma} \label{infsafe}
    Let $P'(M)$ be the precursor to an input-output equation $P(M)$ derived from a rational function model $M$ of the form \cref{eq:model}, assuming the standard ranking described in \cref{ranking}. 
    Let $\tilde{M}$ be the scaled model following the replacement $\tilde{\theta}\mapsto \tilde{\theta}\tilde{u}$ for some $\tilde{\theta} \in \pmb{\theta}$, and let $P(\tilde{M})$ be the resulting corresponding input-output equation to $P(M)$ (i.e. such that $P'(\tilde{M})$ contains the corresponding monomials given in Lemma 1).
    
    Then the parameter solutions for $P(\tilde{M})$ are contained within the parameter solutions for $P(M)$. 
\end{lemma}

\begin{pf}
    Consider $P'(M)$, associated with some input-output equation $P(M)$ derived from model $M$. 
    $P'(M)$ has the general form $0 = P'(M) = \sum_{i = 1}^n C_i m_i$, where the $C_i$ are polynomials in the parameters, the $m_i$ are monomials in the inputs, outputs, and their derivatives, and $m_n$ is the \textit{leading monomial} (noting also that $n$ here is just a positive integer, not the number of state variables in the system).
    
    Each coefficient $C_i$ can itself be viewed as a polynomial in $\tilde{\theta}$ with coefficients in $\pmb{\theta}\setminus\tilde{\theta}$. We will write this as $C_i = c_i + \tilde{c}_i \tilde{\theta}^{p_i}$, where  $\tilde{c}_i \tilde{\theta}^{p_i}$ is the highest degree term in $\tilde{\theta}$ (possibly with $\tilde{c}_i=0$ if there is no $\tilde{\theta}$ present in that $C_i$), and $c_i$ is all remaining terms in $C_i$. $P'(M)$ is then: $0 = P'(M) = \sum_{i = 1}^n (c_i + \tilde{c}_i \tilde{\theta}^{p_i}) m_i = (c_1 + \tilde{c}_1 \tilde{\theta}^{p_1}) m_1 + ... + (c_{n-1} + \tilde{c}_{n-1} \tilde{\theta}^{p_{n-1}}) m_{n-1} + (c_{n} + \tilde{c}_{n} \tilde{\theta}^{p_{n}}) m_{n}$.

    Next, we make $P'(M)$ a monic polynomial $P(M)$ by normalizing with respect to the leading coefficient of $m_n$:
    \begin{fleqn}
    \begin{align*}
        0 = P(M) = &\frac{c_1 + \tilde{c}_1 \tilde{\theta}^{p_1}}{c_{n} + \tilde{c}_{n} \tilde{\theta}^{p_{n}}} m_1 + ... \\&~~~~~+ \frac{c_{n-1} + \tilde{c}_{n-1} \tilde{\theta}^{p_{n-1}}}{c_{n} + \tilde{c}_{n} \tilde{\theta}^{p_{n}}} \tilde{m}_{n-1} + m_n
    \end{align*}
    \end{fleqn}
    which now satisfies the definition of an input-output equation. The parameter solutions of $P(M)$ must therefore satisfy:
    \begin{align}
        \frac{c_i + \tilde{c}_i \tilde{\theta}^{p_i}}{c_n + \tilde{c}_n \tilde{\theta}^{p_n}} & = \frac{c^*_i + \tilde{c}_i^* \tilde{\theta}^{*~p_i}}{c_n^* + \tilde{c}_n^* \tilde{\theta}^{*~p_n}}, ~~~\text{for}~i=1,...,n-1 \label{psoln_thm1}
    \end{align}
    where $c_j^*$ and $\tilde{\theta}^*$ represent the true parameter values.
    
    Similarly, $P'(\tilde{M})$ is a polynomial of at least $n$ monomials now calculated with $\tilde{\theta}$ scaled by a known forcing input function $\tilde{u}$. 
    By Lemma \ref{newform}, $P'(\tilde{M})$ has the general form: 
     \begin{align*}
        0 = P'(\tilde{M}) & =  \sum_{i = 1}^{n} a(c_i + \tilde{c}_i \tilde{\theta}^{p_i}\tilde{u}^{p_i} )  \tilde{m}_i + \sum_{j = 1}^{k} a_j \\
        & =  a(c_1 + \tilde{c}_1 \tilde{\theta}^{p_1} \tilde{u}^{p_1}) \tilde{m}_1 + \cdots \\
        & ~~~~~~~~+ a(c_{n} + \tilde{c}_{n} \tilde{\theta}^{p_{n}}\tilde{u}^{p_{n}}) \tilde{m}_{n} + \cdots
    \end{align*}
    where $\tilde{m}_i$ contain $m_i$ along with any additional monomial of $\tilde{u}$, and $a$ is a multiplicative coefficient in the parameters. The $a_j$ are any additional terms also generated (which by Lemma \ref{newform} are not the leading term).
    
    To make $P'(\tilde{M})$ a monic polynomial $P(\tilde{M})$, we normalize with respect to the coefficient of the leading monomial in $P'(\tilde{M})$. 
    
    If $\tilde{c}_n \neq 0$, then the leading monomial of $P(\tilde{M})$ is $\tilde{u}^{p_n}m_n$, and its coefficient is $a\tilde{c}_n \tilde{\theta}^{p_n}$.
    Then the terms in $P(\tilde{M})$ corresponding to the original monomials $m_i$ (now possibly split) are of the form $\frac{c_i}{\tilde{c}_n \tilde{\theta}^{p_n}} \tilde{m}_i$ and $\frac{\tilde{c}_i \tilde{\theta}^{p_i}}{\tilde{c}_n \tilde{\theta}^{p_n}} \tilde{u}^{p_i} \tilde{m}_i$.
    The parameter solutions of $P(\tilde{M})$ must therefore satisfy:
    \begin{align}
        \frac{c_i}{\tilde{c}_n \tilde{\theta}^{p_n}} & = \frac{c^*_i}{\tilde{c}_n^* \tilde{\theta}^{*~p_n}}, ~~~\text{for}~i=1,...,n \label{pnewsoln_thm1} \\
        \frac{\tilde{c}_i \tilde{\theta}^{p_i}}{\tilde{c}_n \tilde{\theta}^{p_n}} & = \frac{\tilde{c}^*_i \tilde{\theta}^{*~p_i}}{\tilde{c}_n^* \tilde{\theta}^{*~p_n}}, ~~~\text{for}~i=1,...,n \label{pnewsolnB_thm1} 
    \end{align} 
   
    We want to show that the parameter solutions of $P(\tilde{M})$ are also parameter solutions for $P(M).$
    From the parameter solutions \cref{pnewsoln_thm1} and \cref{pnewsolnB_thm1} of $P(\tilde{M})$, we have:
    \begin{align*}
        c_i & = c^*_i \frac{\tilde{c}_n \tilde{\theta}^{p_n}}{\tilde{c}_n^* \tilde{\theta}^{*~p_n}}, & ~~\text{for~}i=1,...,n \\
        \tilde{c}_i \tilde{\theta}^{p_i} & = \tilde{c}_i^* \tilde{\theta}^{*~p_i} \frac{\tilde{c}_n \tilde{\theta}^{p_n}}{\tilde{c}_n^* \tilde{\theta}^{*~p_n}}, & ~~\text{for~}i=1,...,n
    \end{align*}
    Substituting these into the left-hand side of \cref{psoln_thm1}:
    \begin{fleqn}
    \begin{align*}
        \frac{c_i + \tilde{c}_i \tilde{\theta}^{p_i}}{c_n + \tilde{c}_n \tilde{\theta}^{p_n}} &= \frac{c^*_i \frac{\tilde{c}_n \tilde{\theta}^{p_n}}{\tilde{c}_n^* \tilde{\theta}^{*~p_n}} + \tilde{c}_i^* \tilde{\theta}^{*~p_i} \frac{\tilde{c}_n \tilde{\theta}^{p_n}}{\tilde{c}_n^* \tilde{\theta}^{*~p_n}}}{c_n^* \frac{\tilde{c}_n \tilde{\theta}^{p_n}}{\tilde{c}_n^* \tilde{\theta}^{*~p_n}} + \tilde{c}_n^* \tilde{\theta}^{*~p_n} \frac{\tilde{c}_n \tilde{\theta}^{p_n}}{\tilde{c}_n^* \tilde{\theta}^{*~p_n}}} \\
        &= \frac{(c^*_i + \tilde{c}_i^* \tilde{\theta}^{*~p_i}) \frac{\tilde{c}_n \tilde{\theta}^{p_n}}{\tilde{c}_n^* \tilde{\theta}^{*~p_n}}}{(c_n^* + \tilde{c}_n^* \tilde{\theta}^{*~p_n}) \frac{\tilde{c}_n \tilde{\theta}^{p_n}}{\tilde{c}_n^* \tilde{\theta}^{*~p_n}}} = \frac{c^*_i + \tilde{c}_i^* \tilde{\theta}^{*~p_i}}{c_n^* + \tilde{c}_n^* \tilde{\theta}^{*~p_n}}
    \end{align*}
    \end{fleqn}
    This matches the right-hand side of \cref{psoln_thm1}. Therefore the solutions are self-consistent.
    We conclude that solutions of $P(\tilde{M})$ are also parameter solutions for $P(M)$. Note that this calculation becomes slightly more complicated if the $c_i$ include lower powers of $\tilde{\theta}$ (as then the $c_i$ split into distinct monomials), but the basic argument still holds. 
    
    In the case when $\tilde{c}_n = 0$, the coefficient $C_n$ does not contain $\tilde{\theta}$, so after the replacement $\tilde{\theta} \mapsto \tilde{\theta}\tilde{u}$, we have $C_n$ unchanged. The leading term coefficient in both $P(M)$ and $P(\tilde{M})$ is therefore $c_n$ (the highest-degree term in $C_n$ when $\tilde{c}_n = 0$).

After normalization, the parameter solutions must satisfy:
\begin{align*}
\frac{c_i}{c_n} &= \frac{c^*_i}{c_n^*}, \quad \text{and} \quad \frac{\tilde{c}_i \tilde{\theta}^{p_i}}{c_n} = \frac{\tilde{c}_i^* \tilde{\theta}^{*~p_i}}{c_n^*}
\end{align*}
for $\tilde{M}$, and:
\begin{align*}
\frac{c_i + \tilde{c}_i \tilde{\theta}^{p_i}}{c_n + \tilde{c}_n \tilde{\theta}^{p_n}} = \frac{c_i + \tilde{c}_i \tilde{\theta}^{p_i}}{c_n} = \frac{c^*_i + \tilde{c}_i^* \tilde{\theta}^{*~p_i}}{c_n^*}
\end{align*}
for $M$ (since $\tilde{c}_n = 0$). The containment of solution sets follows immediately by adding the two equations from $\tilde{M}$. $\square$ 
\end{pf}

\noindent Lemma~\ref{infsafe} provides assurance that when we scale a parameter with a forcing function, we are not creating new, unexpected parameter solutions that could worsen our identifiability.

This enables us to state Theorem~\ref{nothinglostthm}, which directly addresses the preservation of individual parameter identifiability when forcing functions are introduced.

\begin{theorem}\label{nothinglostthm}
   Let $M$ be a rational function model of the form of \cref{eq:model}, assuming the standard ranking described in \cref{ranking}, and let $\tilde{u}(t)$ be a known, time-dependent forcing input function. 
   Scale any parameter $\tilde{\theta} \in \pmb{\theta}$ with $\tilde{u}(t)$ to generate a new model $\tilde{M}$. 
   Then each parameter that was locally (respectively globally) identifiable in $M$ will be locally (respectively globally) identifiable in $\tilde{M}$.
\end{theorem}
\begin{pf}
    From Lemma~\ref{infsafe}, the solutions to the coefficient map for $\tilde{M}$ will be contained within those of $M$. 
    If a parameter $\theta_k$ is globally identifiable in $M$, it means that for any $\pmb{\theta}'$ such that $c(\pmb{\theta}) = c(\pmb{\theta}')$, we must have $\theta_k = \theta_k'$. Since the solution set for $\tilde{M}$ is a subset of that for $M$, any $\tilde{\pmb{\theta}}'$ yielding $c(\tilde{\pmb{\theta}}) = c(\tilde{\pmb{\theta}}')$ must also satisfy the original relations, implying $\theta_k = \theta_k'$. Thus, $\theta_k$ remains globally identifiable.
    A similar argument holds for local identifiability: if $\theta_k$ has a finite number of solutions in $M$ for a given output, it will have at most that many (and possibly fewer) in $\tilde{M}$.  $\square$
\end{pf}

\noindent \textbf{Significance of Theorem~\ref{nothinglostthm}:} This theorem provides a formal guarantee to modelers that the act of incorporating a known, time-varying input (a forcing function) into a parameter, a common step to better reflect real-world system dynamics or to experimentally probe a system, will not inadvertently destroy or degrade any pre-existing structural identifiability. This safeguard allows researchers to confidently employ forcing functions as a tool, particularly when aiming to improve identifiability, as explored in \Cref{s:forcefunc}.

\begin{corollary}\label{cor1}
   Let $M$ be a rational function model of the form of \cref{eq:model}, assuming the standard ranking described in \cref{ranking}, and let $\tilde{u}(t)$ be a known, time-dependent forcing input function.
   Scale any parameter $\tilde{\theta} \in \pmb{\theta}$ with $\tilde{u}(t)$ to generate a new model $\tilde{M}$. If $M$ is locally (respectively globally) identifiable, then $\tilde{M}$ is as well.
\end{corollary}
\begin{pf}
    This follows directly from Theorem~\ref{nothinglostthm}. 
    If all parameters in the model $M$ are locally (globally) identifiable, and each remains so in $\tilde{M}$, then $\tilde{M}$ will also be locally (globally) identifiable. $\square$
\end{pf}

\begin{corollary} \label{cor2}
   Let $M$ be a rational function model of the form of \cref{eq:model}, assuming the standard ranking described in \cref{ranking}, and let $\tilde{u}(t)$ be a known, time-dependent forcing input function.
   Replace any parameter $\tilde{\theta} \in \pmb{\theta}$ with $\tilde{u}(t)$ to generate a new model $\tilde{M}$. If $M$ is locally (respectively globally) identifiable, then $\tilde{M}$ is as well.
\end{corollary}
\begin{pf}
    Observe that replacement is a special case of Corollary~\ref{cor1} where we additionally fix $\tilde{\theta} = 1$. Since fixing a parameter value generically will not worsen identifiability of the remaining parameters (it can only add constraints or reduce the parameter space), the result holds. $\square$
\end{pf}

\noindent \textbf{Implication of Corollaries~\ref{cor1} and \ref{cor2}:} These corollaries extend the assurance of Theorem~\ref{nothinglostthm} to the entire model. If a model is already identifiable, introducing a forcing function (either by scaling an existing parameter or replacing it) will not render the model unidentifiable. This provides confidence when modifying well-characterized models with new dynamic inputs.

While the previous section established that forcing functions do not harm identifiability, we now demonstrate when they actively improve it. The following theorems delineate specific conditions for achieving this improvement.

\section{Forcing inputs can improve identifiability} \label{s:forcefunc}
\noindent Having established that forcing functions do not harm identifiability (\Cref{s:incorp}), this section presents our second main contribution: theorems that delineate specific conditions under which these inputs actively \textit{improve} it, potentially resolving unidentifiability or transitioning local to global identifiability. 

We will systematically explore scenarios starting with common identifiable combinations involving one degree of freedom, first addressing additive structures through scaling (Theorem~\ref{scaling_1d}, \Cref{s:scaling_1dof}) and then considering more general combinations through parameter replacement (Theorem~\ref{replace_1d_FINAL}, \Cref{Sss:UnidentNonlinReplace_forcefunc}). We then extend this to more complex entanglements involving more than one degree of freedom (Theorem~\ref{replace_2d}, \Cref{s:replace_multi_dof}). Throughout, corollaries will highlight the implications for the overall model's identifiability. Finally, we will connect these analytical results to the visual intuition provided by parameter graphs (\Cref{s:connecting_param_graph_forcefunc}), demonstrating how these theorems can be practically guided. 

To do this, we restrict our analysis to scenarios where all parameters appear in the input-output equation, meaning there are no parameters that are entirely insensitive to the observed output.

\subsection{Scaling under restrictions to correct for 1 degree of freedom in parameter solution set} \label{s:scaling_1dof}
\noindent We begin by considering the simplest case of unidentifiability: an additive combination of two parameters. Theorem~\ref{scaling_1d} demonstrates how scaling one of these parameters with a forcing function can disentangle them.

\begin{theorem} \label{scaling_1d}
    Let $M$ be any structurally unidentifiable rational function model of the form of \cref{eq:model}, assuming the standard ranking described in \cref{ranking}, and suppose that $P(M)$ contains a parameter $\tilde{\theta} \in \pmb{\theta}$ in an additive combination with a rational function $f$ of one other parameter $\theta_i \in \pmb{\theta} \setminus \{\tilde{\theta}\}$ such that neither $\tilde{\theta}$ nor $\theta_i$ is identifiable but the combination $(f(\theta_i) + \tilde{\theta})$ is locally (resp. globally) identifiable. Suppose $(f(\theta_i) + \tilde{\theta})$ appears as a  multiplicative part of at least one coefficient in an input-output equation $P(M)$, such that it may be multiplied by some rational function of the parameters.
 
    Scale $\tilde{\theta}$ with a known forcing input function $\tilde{u}(t)$ to generate a new model $\tilde{M}$. 
    Then $\tilde{\theta}$ is locally (resp. globally) identifiable in $P(\tilde{M})$. Additionally, if $f$ is injective, then $\theta_i$ is locally (resp. globally) identifiable as well.
\end{theorem}

\begin{pf}
    Suppose that $P(M)$ contains an identifiable combination of the form described, $(f(\theta_i) + \tilde{\theta})$, where $f$ is a rational function of $\theta_i \in \pmb{\theta} \setminus \{\tilde{\theta}\}$ (as an aside, we note that $f$ may also be a function of other identifiable parameters as well and this will not affect the proof). 
    
    Also note that from the theorem statement there exists at least one term of the form $a(f(\theta_i) + \tilde{\theta})m$ in $P(M)$, where $a$ is a rational function of the parameters and $m$ is a monomial in the inputs, outputs, and their derivatives. Since input-output equation coefficients are globally identifiable, and $(f(\theta_i) + \tilde{\theta})$ is locally (globally) identifiable, $a$ must also be locally (globally) identifiable.

    By Lemma \ref{infsafe}, since $a$ and $(f(\theta_i) + \tilde{\theta})$ are locally (resp. globally) identifiable in $M$, they must be so in $\tilde{M}$ (since incorporating the forcing function cannot add any new parameter solutions). Also note that by Lemma \ref{newform}, when $\tilde{u}$ scales $\tilde{\theta}$ to form the model $\tilde{M}$, the original monomial term $a (f(\theta_i) + \tilde{\theta}) m$ will transform in the new input-output relation $P(\tilde{M})$ to $ab f(\theta_i) \tilde{m} + ab\tilde{\theta}\tilde{u}(t) \tilde{m}$, where $b$ is a rational function of the parameters that results from a potentially different normalizing coefficient in $P(\tilde{M})$, and $\tilde{m}$ is a monomial containing $m$ as well as potentially some powers of $\tilde{u}$. (Noting that the combination could not have been part of the leading term of $P'(M)$ as it appears in the numerator of $P(M)$ and so will appear in $P'(\tilde{M})$ and then $P(\tilde{M})$ in this form by Lemma 1.) 
    
    Now $ab f(\theta_i) \tilde{m}$ and $ab\tilde{\theta}\tilde{u}(t) \tilde{m}$ are distinct monomial terms, making their coefficients ($ab f(\theta_i)$ and $ab\tilde{\theta}$) separately identifiable. Since $(f(\theta_i) + \tilde{\theta})$, $a$, $ab f(\theta_i)$, and $ab\tilde{\theta}$ are each identifiable, we can solve uniquely for $f(\theta_i)$, $\tilde{\theta}$, and $b$, making $\tilde{\theta}$ locally (globally) identifiable. Additionally, if $f$ is injective then $\theta_i$ is locally (resp. globally) identifiable.  $\square$ 
\end{pf}

\noindent \textit{Remark.}
This theorem applies when the additive combination appears as a multiplicative part of the coefficient of a monomial term in $P(M)$. 
However, a very similar argument will hold for a range of more general forms of the coefficient containing $(f(\theta_i) + \tilde{\theta})$ (e.g. when the additive term appears in the denominator, or appears with a more general form in the numerator, etc.). 
Extension of this theorem to such cases is straightforward but requires case-by-case verification.
%}

\begin{corollary}\label{cor3}
   Let \cref{scaling_1d} be satisfied with $f$ injective, and suppose $(f(\theta_i) + \tilde{\theta})$ and the parameters in $\pmb{\theta} \setminus \{\theta_i, \tilde{\theta} \}$ are locally (respectively globally) identifiable in $M$.
   Then $\tilde{M}$ must be locally (respectively globally) identifiable. 
\end{corollary}
\begin{pf}
    By assumption, all parameters in $\pmb{\theta} \setminus \{\theta_i, \tilde{\theta} \}$ are locally (respectively globally) identifiable in $M$, and by Theorem~\ref{nothinglostthm}, they remain so in $\tilde{M}$.
    
    From Theorem~\ref{scaling_1d}, after scaling,  $\tilde{\theta}$ and $\theta_i$ become locally (globally) identifiable in $\tilde{M}$.
    Since all parameters in the model $\tilde{M}$ are now locally (globally) identifiable, then $\tilde{M}$ itself is locally (globally) identifiable. $\square$
\end{pf}

\noindent \textbf{Significance of Theorem~\ref{scaling_1d} and Corollary~\ref{cor3}:} These results provide a direct method to resolve unidentifiability in common additive parameter structures. By strategically scaling one parameter involved in such a sum with a known time-varying input, both parameters can become globally identifiable, potentially rendering an entire previously unidentifiable model identifiable. This is particularly useful when direct measurement of one parameter is difficult, but a correlated, measurable forcing function can be applied. This aligns with recent work on multi-experiment identifiability theory for ODE systems with time-varying coefficients, showing how multiple experiments with different inputs can recover identifiability \cite{ovchinnikov2022multiexperiment}.

\subsection{Replacement to correct for 1 degree of freedom in parameter solution set} \label{Sss:UnidentNonlinReplace_forcefunc}
\noindent While scaling is effective for additive combinations, parameter replacement offers a more general approach for other types of two-parameter entanglements. Theorem~\ref{replace_1d_FINAL} shows how replacing one parameter in such a combination with a known forcing function can make the other identifiable.

\begin{theorem} \label{replace_1d_FINAL}
    Let $M$ be any structurally unidentifiable rational function model of the form of \cref{eq:model}, assuming the standard ranking described in \cref{ranking}, and suppose that $P(M)$ contains a parameter $\tilde{\theta} \in \pmb{\theta}$ in an identifiable combination with one other parameter $\theta_i \in \pmb{\theta}\setminus \tilde{\theta}$ such that neither is separately identifiable but the locally (resp. gloablly) identifiable combination $g$ is injective for a fixed $\tilde{\theta}$ (e.g. $g = \tilde{\theta}\theta_i$). %, and suppose that this combination appears as the coefficient of a monomial term in $P(M)$ (i.e., of the form $g(\tilde{\theta}, \theta_i)m$ where $m$ is a monomial in outputs, inputs, and their derivatives).
 
    Replace $\tilde{\theta}$ with a known forcing input function $\tilde{u}$ to generate a new model $\tilde{M}$. 
    Then $\theta_i$ must be locally (resp. globally) identifiable in $P(\tilde{M})$.
\end{theorem}
\begin{pf}
    Suppose that $g(\tilde{\theta}, \theta_i)$ is a function representing the locally (respectively globally) identifiable combination with one other parameter $\theta_i$ from $\pmb{\theta}\setminus \tilde{\theta}$, such that $g$ is injective for a fixed $\tilde{\theta}$.

    If we scale (but not yet replace) $\tilde{\theta}$ by $\tilde{u}$, we know from Lemma \ref{infsafe} that $g(\tilde{\theta}, \theta_i)$ must remain identifiable, as scaling with a forcing function cannot introduce new parameter solutions. Then to consider replacement of $\tilde{\theta}$ by $\tilde{u}$, we now set $\tilde{\theta}=1$ from our scaled model, yielding that $\theta_i$ must be locally (respectively globally) identifiable as $g$ is injective for a fixed $\tilde{\theta}$. As an aside, note that if $g$ is not injective but has finite solutions (e.g. $\tilde{\theta}\theta_i^2$), then $\theta_i$ is similarly locally identifiable. $\square$ 
\end{pf}

\begin{corollary}\label{cor5}
   Let \cref{replace_1d_FINAL} be satisfied and restrict $g$ and the parameters in $\pmb{\theta} \setminus \{\theta_i, \tilde{\theta} \}$ to be locally (respectively globally) identifiable.
   Then $\tilde{M}$ must be locally (respectively globally) identifiable.
\end{corollary}
\begin{pf}
    By assumption, all parameters in $\pmb{\theta} \setminus \{\theta_i, \tilde{\theta} \}$ are locally (respectively globally) identifiable in $M$, and by Theorem~\ref{nothinglostthm} (adapted for replacement), they remain so in $\tilde{M}$.
    From Theorem~\ref{replace_1d_FINAL}, after replacement, $\theta_i$ is locally (globally) identifiable in $\tilde{M}$, and $\tilde{M}$ no longer contains $\tilde{\theta}$ as an unknown parameter.
    Since each remaining parameter in the model $\tilde{M}$ is locally (globally) identifiable, then $\tilde{M}$ itself is locally (globally) identifiable. $\square$
\end{pf}
\noindent \textbf{Significance of Theorem~\ref{replace_1d_FINAL} and Corollary~\ref{cor5}:} These results are important because they broaden the applicability of forcing functions. By replacing an unidentifiable parameter with a known input, we effectively reduce the dimensionality of the unknown parameter space. This can resolve the identifiability of its entangled partner and, by extension, the entire model.

Equivalently, by replacing an unknown parameter with a known function, we have reduced the parameter space by 1, and therefore corrected for at least 1 degree of freedom in the parameter solution set space. 
Below we prove a case where replacing an unknown parameter with a known function reduces the degrees of freedom in the parameter solution set space by more than one.

\subsection{Replacement to correct for more than 1 degree of freedom in parameter solution set} \label{s:replace_multi_dof}
\noindent We now consider a more complex scenario as an example where unidentifiability stems from a combination involving three parameters in both additive and multiplicative relationships. Theorem~\ref{replace_2d} illustrates how replacing one strategically chosen parameter can resolve the identifiability of the other two.

\begin{theorem}\label{replace_2d}
    Let $M$ be any structurally unidentifiable model of the form of \cref{eq:model}, assuming the standard ranking described in \cref{ranking}, and suppose the coefficient map $c(\pmb{\theta})$ of $P(M)$ contains a term of the form $\theta_j (\tilde{\theta} + \theta_i)$ where $\theta_i, ~\theta_j \in  \pmb{\theta} \setminus \{\tilde{\theta}\}$, $i \neq j$. Let $\theta_i$, $\theta_j$, and $\tilde{\theta}$ all be unidentifiable parameters in $M$.
 
    Replace $\tilde{\theta}$ with a known non-constant forcing input function $\tilde{u}$ to generate a new model $\tilde{M}$. 
    Then $\theta_i$ and $\theta_j$ must be globally identifiable in $P(\tilde{M})$.
\end{theorem}
\begin{pf}
    Let $a_m$ be a term in $c(\pmb{\theta})$ of $P(M)$ and have the general form $a_m = \theta_j (\tilde{\theta} + \theta_i)$, where $\theta_i, ~\theta_j \in  \pmb{\theta} \setminus \{\tilde{\theta}\}$, $i \neq j$. 
    Then in the input-output relation of $M$ this corresponds to the term $\theta_j (\tilde{\theta} + \theta_i)m$, where $m$ is a monomial in the $y$ and $u$ variables (and their derivatives).

    Let us write $\theta_j (\tilde{\theta} + \theta_i) = \theta_j \tilde{\theta} + \theta_j \theta_i$ and begin by scaling $\tilde{\theta}$ with $\tilde{u}$ (rather than replacing). Then by a very similar argument to Theorem  \ref{scaling_1d}, we can see that after scaling, $\theta_j \tilde{\theta}$ and $\theta_j \theta_i$ are each separately globally identifiable.

    Now, to complete the replacement step, we set $\tilde{\theta} = 1$, which yields that $\theta_j$ alone is now globally identifiable, which from the identifiable combination $\theta_j \theta_i$ gives that $\theta_i$ is also globally identifiable. $\square$
\end{pf}

\noindent
\textit{Remark.} We note that, similar to Theorem \ref{scaling_1d}, this theorem can be extended relatively straightforwardly to cases where the identifiable combination appears in the input-output equation in other ways rather than purely as the coefficient of a monomial term (e.g. multiplied by another rational function of the parameters, etc.). \medskip

\noindent
\textit{Remark.} Additionally, the alternate extension $\tilde{\theta}(\theta_i + \theta_j)$ requires additional consideration. Under the assumption that $\theta_i$, $\theta_j$, and $\tilde{\theta}$ are unidentifiable parameters, replacing $\tilde{\theta}$ with $\tilde{u}(t)$ yields $\tilde{u}(t)(\theta_i + \theta_j)$. This makes the sum $\theta_i + \theta_j$ identifiable, but does \textit{not} resolve the individual identifiability of $\theta_i$ and $\theta_j$ without further information or assumptions.

\begin{corollary}\label{cor6}
   Let \cref{replace_2d} be satisfied and suppose that the parameters in $\pmb{\theta} \setminus \{\theta_i, \theta_j, \tilde{\theta} \}$ be locally (respectively globally) identifiable in $M$.
   Then $\tilde{M}$ must be locally (respectively globally) identifiable.
\end{corollary}
\begin{pf}
    By assumption, all parameters in $\pmb{\theta} \setminus \{\theta_i, \theta_j, \tilde{\theta} \}$ are locally (respectively globally) identifiable in $M$ and remain so in $\tilde{M}$.
    From Theorem~\ref{replace_2d}, after replacement, $\theta_i$ and $\theta_j$ are globally identifiable in $\tilde{M}$, and $\tilde{M}$ no longer contains $\tilde{\theta}$ as an unknown.
    Since all remaining parameters in the model $\tilde{M}$ are locally (globally) identifiable, then $\tilde{M}$ itself is locally (globally) identifiable. $\square$
\end{pf}
\noindent \textbf{Significance of Theorem~\ref{replace_2d} and Corollary~\ref{cor6}:} The power of these results is in demonstrating that a single forcing input (via replacement) can resolve more complex, multi-parameter unidentifiabilities. This highlights the potential for targeted interventions to have a cascading effect on improving model identifiability, especially in systems where parameters are entangled in mixed additive and multiplicative forms.

\subsection{Connecting forcing functions and the parameter graph} \label{s:connecting_param_graph_forcefunc}
\noindent By examining the parameter graph for a particular ODE system, we can build intuition on how to identify more complicated connected components that would separate into identifiable parameters (therefore correcting for more than 1 degree of freedom in the parameter set solution) with only one parameter being replaced.

\begin{figure}
    \centering
    \includegraphics[height = 1.85cm]{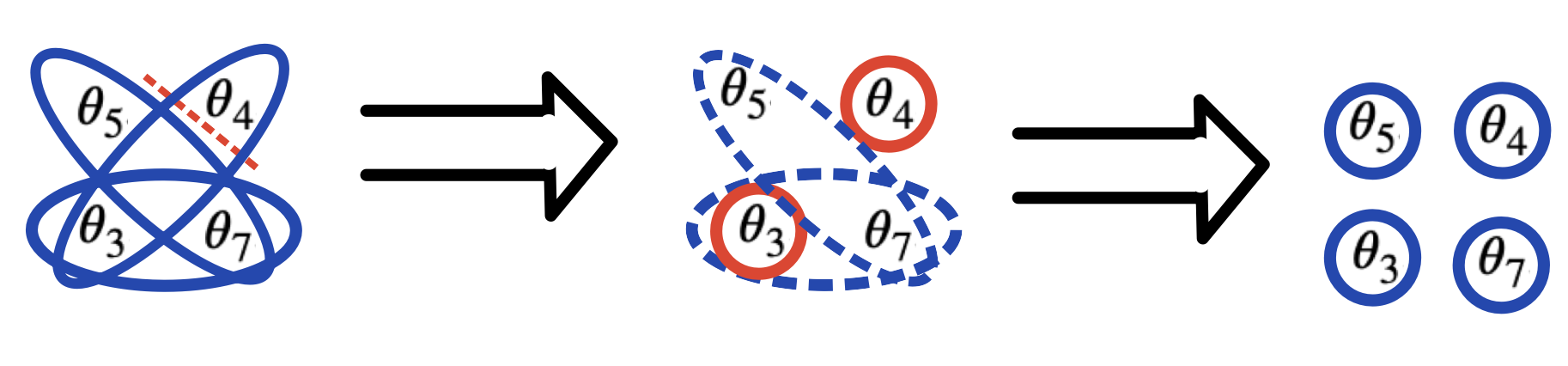}
    \caption{Step by step visualization of replacing $\theta_4$ in the parameter graph from \cref{fig:bilin_paramgraph}(b). Replacing $\theta_4$ with a known forcing input function (or even a known constant) would make the other parameters in the connected component structurally identifiable. These visualizations guide researchers on how to replace parameters to make the system structurally identifiable. Scaling will require more knowledge of the exact parameter relationship(s), as was explored in \Cref{s:incorp}.}
    \label{fig:howto}
\end{figure}

For example, consider the connected component in \cref{fig:howto}.
By replacing the unknown and unidentifiable $\theta_4$ with a known data stream (or known constant), the other parameters in the connected component become identifiable.
In \Cref{ss:bilin_appendix}, we further explain which compartmental models can give rise to this exact connected component.

\noindent In this way, the parameter graph for any rational function ODE system can be used to guide decisions on which subset of parameters to measure to improve structural identifiability.

\section{Procedure in Practice} \label{s:method}
\noindent This section outlines a general framework for applying the theorems from \Cref{s:incorp} and \Cref{s:forcefunc} to practical modeling scenarios. We first detail a systematic process for translating the theory into actionable steps for model formulation and identifiability analysis when initial checks reveal unidentifiability (\Cref{s:translating_theory}). The subsequent section (\Cref{s:generic_ex_main}) will then provide concrete examples using generic linear models.

\subsection{Translating theory to practice} \label{s:translating_theory}
\noindent We provide an explicit framework for taking the theory proved in \Cref{s:forcefunc} to application.
Importantly, this process assumes the existence of parameters that could reasonably be scaled or replaced with a forcing function, either by incorporating more data (e.g. climate drivers) or by experimentally perturbing/forcing particular parameters of interest.

Given a problem of interest, first define a rational function ODE model $M(\pmb{\theta})$ that describes the system of interest with measurable output(s) $\pmb{y}$.
Then, check the structural identifiability of $M(\pmb{\theta})$ using an algebraic software such as DAISY \cite{bellu2007daisy}, SIAN \cite{hong2019sian}, StructuralIdentifiability \cite{dong2023differential}, or STRIKE-GOLDD \cite{diaz2023strike}.
Example input files for DAISY can be found in \cref{DAISYex}.
For small systems, this analysis can possibly be done by hand.
If the system is structurally identifiable, proceed with parameter estimation or prediction modeling. If not, consider the process flow below:

\begin{mynote}{Model Formulation Steps}{n_subprocess}
        %\noindent Check the identifiability of $M:$
        \begin{enumerate}
            \item Determine which parameters are in parameter combinations and generate a parameter graph.
            \item For each connected component, identify which parameter(s) could make the component identifiable. 

            \noindent Consider the following:
            \begin{itemize}
                \item Can the parameter be scaled with a forcing function input? Then scale the chosen parameter with a measured input stream to generate $\Tilde{M}$.

                \item Replace a chosen parameter from a connected component with a measured input to generate $\Tilde{M}$. 
                \item Re-parameterize by replacing a chosen parameter combination with a single new parameter to generate $\Tilde{M}$. 
            \end{itemize}  
        \end{enumerate}
    \label{n_subprocess}
\end{mynote}

\section{Generic Linear Model and how to Introduce Forcing Functions} \label{s:generic_ex_main}

This section provides a concrete illustration of applying forcing functions to improve identifiability in a common model structure. We focus on a generic 2-dimensional linear model, demonstrating through its input-output relation how the introduction of forcing functions via scaling or replacement can alter parameter identifiability. This serves as a practical example of the principles discussed in \Cref{s:method} and the theorems from \Cref{s:incorp} and \Cref{s:forcefunc}. More extensive analysis of linear and bilinear models up to 3 dimensions can be found in \Cref{s:3dim_appendix}.

For a generic linear model, we take advantage of the $\dot{\pmb{x}} = A\pmb{x} + \pmb{b}u$ structure, initially generating a model with movement into/out of all compartments by all compartments.
To constrain and simplify the model, we impose $\pmb{b} = 0$, such that there are no existing inputs into the system.
Finally, we constrain the measured component to only one state variable, of the form $y =x_1$.

Examples of linear models can be found in the literature, beginning with the pharmacokinetics model that we showed in \Cref{ss:MotEx2_diff_alg}. 
Other examples include but are not limited to damped harmonic oscillators, radiation decay chains, or circuit design \cite{belyshev2014studying, bauer2001transport,chandrasekar2007lagrangian, dekker1981classical, fu2021solving, bellen1999methods, lohfink2005linear}.

Here we explore one illustrative structural identifiability problem for a linear system of dimension 2.
Consider a system with two state equations, with the general form:
\begin{align} \label{eq:2da_generic}
    \dot{x}_1 &= \theta_1 x_1 + \theta_2 x_2 \\ \nonumber
    \dot{x}_2 &= \theta_3 x_1 + \theta_4 x_2 \\ \nonumber
    y &= x_1
\end{align}
assuming that only one state at most is measured ($y$).
Then the input-output relation will be:
\begin{align}\label{eq:io_2d_generic_}
        0 = \ddot{y} - (\theta_1 + \theta_4) \dot{y} + (\theta_1 \theta_4 - \theta_2 \theta_3) y
\end{align} 
This makes the coefficient map the following: $c(\pmb{\theta}) = \{\theta_1 + \theta_4, \theta_1 \theta_4 - \theta_2 \theta_3\}$.
This relation is unidentifiable if all parameters are distinct and nonzero.

\begin{figure}
    \centering
    \includegraphics[width=0.5\textwidth]{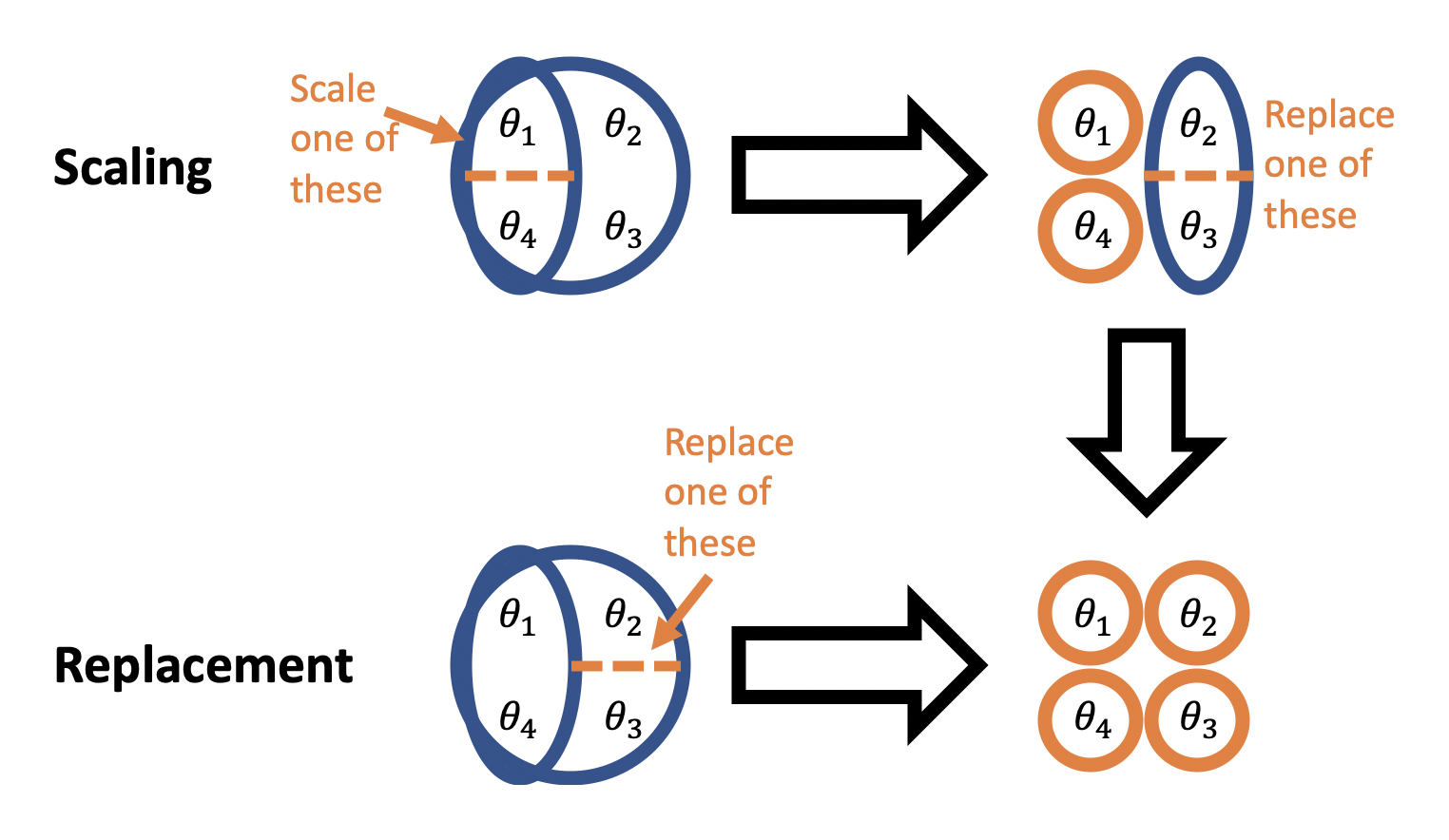}
    \caption{Parameter graph of unidentifiable parameters for the 2D Linear model when all parameters are nonzero. All parameters shown have a functional relationship with at least one other parameter, and cannot be separated as they are currently, but can be resolved through the use of a scaling and/or replacement. Top row: scaling $\theta_1$ or $\theta_4$ improves the identifiability but is not sufficient to fully resolve it for all parameters. Bottom row: replacing $\theta_2$ or $\theta_3$ on the other hand will fully resolve the identifiability of the system (either on its own or as a follow-up step to the previous scaling). }
    \label{fig:LinearCase2_generic}
\end{figure}

While the identifiable combinations of this system cannot be immediately resolved to yield global identifiability, we can do so with two alternative applications of the methods presented here (scaling or replacing two parameters with inputs), as illustrated in \Cref{fig:LinearCase2_generic}. 

Observe that $\theta_1$ and $\theta_4$ are in both an additive and a multiplicative relationship, so we may be able to apply the scaling theorem.
On the other hand, $\theta_2$ and $\theta_3$ are in a multiplicative relationship, which violates conditions of the scaling theorem, \Cref{scaling_1d}. Instead, we will test out replacing one with a known input.

\vspace{0.2cm}

\noindent \textbf{Scaling.} First, consider how the input-output relation changes if we scale $\theta_1$ by a known measured input $u(t)$.
\begin{align} \label{eq:2db_generic}
    \dot{x}_1 &= \theta_1 \mathcolorbox{Yellow}{u(t)} x_1 + \theta_2 x_2 \\ \nonumber
    \dot{x}_2 &= \theta_3 x_1 + \theta_4 x_2 \\ \nonumber
    y &= x_1
\end{align}
We then re-calculate the input-output equation for this new system, yielding:
\begin{fleqn}
\begin{align}\label{eq:io_2d_augA_generic}
        0 = \ddot{y} - \theta_1 \mathcolorbox{Yellow}{u(t)} \dot{y} - \theta_4 \dot{y} + \theta_1 \mathcolorbox{Yellow}{\dot{u}(t)} y + \theta_1 \theta_4 \mathcolorbox{Yellow}{u(t)} y - \theta_2 \theta_3 y
\end{align} 
\end{fleqn}
Now, the coefficient map is the following: $c(\pmb{\theta}) = \{\theta_1,  - \theta_4, $

\noindent $\theta_1 \theta_4, - \theta_2 \theta_3\}$.
By scaling $\theta_1$ by $u(t)$, our augmented input-output relation now contains 6 terms, compared to the original 3 terms in \Cref{eq:io_2d_generic_}. 
These new terms allow us to gain information about the original system.
Now, $\theta_1$ and $\theta_4$ are globally identifiable for this generic 2D linear system.
However, $\theta_2$ and $\theta_3$ are still unidentifiable.

\vspace{0.2cm}

\noindent \textbf{Replacement.} Next, we instead consider how the input-output relation changes if we replace $\theta_2$ with a known observation $\tilde{u}(t)$, without scaling $\theta_1$ or $\theta_4$ (although this step could also be performed after such a scaling if desired, with similar results). 
\begin{align} \label{eq:2dc_generic}
    \dot{x}_1 &= \theta_1 x_1 + \mathcolorbox{PowderBlue}{\tilde{u}(t)} x_2 \\ \nonumber
    \dot{x}_2 &= \theta_3 x_1 + \theta_4 x_2 \\ \nonumber
    y &= x_1
\end{align}
Then the new input-output relation becomes:
\begin{multline}
\label{eq:io_2d_augB_generic}
        0 = \mathcolorbox{PowderBlue}{\tilde{u}(t)}\ddot{y} - (\theta_1 + \theta_4) \mathcolorbox{PowderBlue}{\tilde{u}(t)} \dot{y}  - \mathcolorbox{PowderBlue}{\dot{\tilde{u}}(t)}\dot{y} \\
        +  \theta_1 \mathcolorbox{PowderBlue}{\dot{\tilde{u}}(t)} y +  \theta_1 \theta_4 y - \theta_3 \mathcolorbox{PowderBlue}{\tilde{u}(t)^2} y
\end{multline} 
This coefficient map is: $c(\pmb{\theta}) = \{\theta_1 + \theta_4, -1, \theta_1, \theta_1 \theta_4, - \theta_3\}$.
By replacing $\theta_2$ with $\tilde{u}$, we have removed one of the unidentifiable parameters by replacing it with a known function, and new information has been gained beyond what was introduced.
Our system is now globally identifiable with this one forcing function introduction.

Importantly, just assuming that $\theta_2$ is known (but not a time-varying forcing function) would result in an analogous input-output equation to \Cref{eq:io_2d_generic_}, which would only resolve the $\theta_2$ and $\theta_3$ pair rather than resolving the full global identifiability of the system, illustrating the ways that a non-constant parameter can be informative. This can also be observed in part by noting that $\theta_1$ is the coefficient of $\dot{\tilde{u}}(t)$, which would be zero if $\tilde{u}(t)$ was a constant.

However, what this generic linear model example lacks is context. What we will see in the following section is that the context around the problem is also important. We must consider (a) what additional data streams are actually available to use as forcing input functions, and further (b) whether the augmented systems are still viable representations of the original problem.

\section{Examples from the Literature} \label{s:examples}
\noindent To understand this process in more detail, we review two examples from the literature that either (a) already employ the theorems outlined in this paper or (b) could benefit from the proposed strategies for resolving structural identifiability. Specifically, we revisit the pharmacokinetics model previously introduced (\Cref{ss:revisit}) and examine a cholera SIWR model (\Cref{ss:cholera}). These examples illustrate how the theoretical framework can be applied to concrete biological systems. For further examples of applying the Ritt's Pseudo-Division process from \Cref{s:diff_alg_method}, see \Cref{RPDexample}.

\subsection{Revisiting the Pharmacokinetics model} \label{ss:revisit}
\noindent Recall in \Cref{ss:MotEx2_diff_alg} the common pharmacokinetics model described by \cref{ex2_diff_alg}.
We have established that this model is unidentifiable both analytically and visually using the parameter graph in \cref{fig:paramgraph}.
The re-parameterization of this model as proposed by \citet{meshkat2009algorithm} poses issues with interpretability of parameter estimates, and may fail to answer key questions of interest.

We propose applying Theorem~\ref{scaling_1d} to resolve the identifiability issues of this model.
Specifically, recall from the discussion of \cref{fig:paramgraph} that we established only one parameter needs to be measured to break the connected component.
We also already know the analytical form of the coefficient map for \cref{ex2_diff_alg} is given by: $$c(\pmb{\theta}) = \{V, ~k_{12}k_{21}, ~\mathcolorbox{MistyRose}{k_{02}} + k_{12}, ~\mathcolorbox{MistyRose}{k_{01}} + k_{21}\}$$
For the simplest additive relationship with one other parameter condition from Theorem~\ref{scaling_1d}, we can choose to scale either $k_{01}$ or $k_{02}$ ($k_{12}$ or $k_{21}$ would also work, but these are slightly more complex as they are also in a multiplicative combination).

\vspace{0.2cm}
\noindent \textit{Scaling or replacing $k_{02}$:} Recall that $k_{02}$ represents the drug decay rate in tissue.
This may not be directly measurable, as repeatedly taking tissue samples could be painful or even impractical to implement.
In this case, replacing $k_{02}$ with a measured data input is not feasible.

The practical solution is to measure a related process that is known to influence the drug's decay rate in tissue, and use this as a proxy-informant to \textit{scale} $k_{02}$.
This is a reasonable assumption as drug decay rates can change over time.
We can theorize that such a process exists, and applying Theorem~\ref{scaling_1d}, scale $k_{02}$ with a known forcing input function $\tilde{u}(t)$.
DAISY output finds that this small change to the model makes it structurally identifiable.
A brief search of the literature did not reveal whether this theoretical process exists, although potentially either imaging studies with a labeled drug or measurement of a metabolite could be (potentially expensive) options.

%\textcolor{green}{DAISY output for replacing  k02 with a "known" (arbitrarily set to 3 for this example (GLOBALLY IDENTIFIABLE): $$- df(u,t) + df(y,t,2)*v + %df(y,t)*v*(k01 + k12 + 3) + u*( - k12 + k21 - 3) + y*v*(k01*k12 - k01*k21
 %+ 3*k01 + k12*k21 - k21**2 + 3*k21)$$
 %And scaling k02 with a known input u1 (GLOBALLY IDENTIFIABLE): $$- df(u,t) + df(y,t,2)*v + df(y,t)*u1*k02*v + df(y,t)*v*(k01 + k12) - u*u1*k02 + u*( - k12 + k21) + 
%u1*y*k02*v*(k01 + k21) + y*v*(k01*k12 - k01*k21 + k12*k21 - k21**2)$$}

\vspace{0.2cm}
\noindent \textit{Scaling or replacing $k_{01}$:} This parameter represents the drug elimination or decay rate in blood.
In this case, regular blood samples may be taken and the parameter measured directly.
Utilizing such measurement data would be an example of \textit{replacement}, as described in Theorem~\ref{replace_1d_FINAL}.
DAISY output confirmed that replacing $k_{01}$ with a known forcing input function $\tilde{u}(t)$ resolves the structural identifiability issues of the original model.

If the drug decay rate in blood is too low to directly measure, we can instead explore \textit{scaling} $k_{01}$.
It is known, for example, that drug decay in blood depends on pH, which is measurable and changes over time \cite{onetto2023drug}.
As expected from Theorem~\ref{scaling_1d}, scaling $k_{01}$ with a known forcing input function $\tilde{u}(t)$ resolves the structural identifiability issues of the original model, confirmed by DAISY.
\vspace{0.2cm}

\noindent Both of the suggestions for targeting $k_{01}$ have the advantage of (a) maintaining the original parameter interpretations, and (b) being relatively simple to implement in practice.

\subsection{Cholera SIWR Model} \label{ss:cholera}
\noindent The following example illustrates an application of Corollary~\ref{cor1} from the literature, in which a forcing input function was introduced into a nonlinear ODE system without altering the structural identifiability.

\citet{tienMultipleTransmissionPathways2010} present an SIWR model of cholera infection, the non-dimensionalized version of which is shown in \cref{siwr}.
Here $s$ is the proportion of susceptible individuals in the population, $i$ is the proportion of infected individuals, and $r$ is the recovered or removed proportion of individuals.
The $w$ compartment represents the concentration of cholera in the water, one of the main transmission pathways for cholera infection.
$\beta_W$ is the transmission parameter for waterborne cholera transmission, while $\beta_I$ is the parameter for direct transmission.
$\gamma$ is the recovery rate and $\xi$ is the pathogen decay rate in water.
\begin{align}
    \dot{s} & = -\beta_W s w - \beta_I s i \nonumber \\
    \dot{i} & = \beta_W s w + \beta_I s i - \gamma i \nonumber \\
    \dot{w} & = \xi(i - w) \nonumber \\
    \dot{r} & = \gamma i \label{siwr} \\
    y & = k i \nonumber
\end{align}
\citet{eisenberg2013identifiability} proved that this model is structurally identifiable given the measured output of prevalence $ k_i$, where $k$ is the reporting rate of cholera cases.
We confirmed these results in DAISY.

\citet{eisenberg2013examining} attempt to use the above model to fit cholera data from a 2010 outbreak in Haiti.
While \cref{siwr} was shown to be structurally identifiable, they find that there are \textit{practical} identifiability issues with fitting the available data to the model.
\vspace{0.2cm}

\noindent \textit{Scaling $\beta_W$:} Heavy rainfall during this outbreak was known to spur waterborne transmission of cholera across Haiti.
\citet{eisenberg2013examining} proposed scaling the waterborne transmission parameter $\beta_W$ with rainfall data $f(t)$:
\begin{align}
    \dot{s} & = -\mathcolorbox{PowderBlue}{f(t)} \beta_W  s w - \beta_I s i \nonumber \\
    \dot{i} & = \mathcolorbox{PowderBlue}{f(t)} \beta_W  s w + \beta_I s i - \gamma i \nonumber \\
    \dot{w} & = \xi(i - w) \nonumber \\
    \dot{r} & = \gamma i \label{siwr2} \\
    y & = k i \nonumber
\end{align}
\citet{eisenberg2013examining} show that \cref{siwr2} resolved the practical identifiability issue, but did not explicitly show that \cref{siwr2} maintained the structural identifiability of \cref{siwr}.

We argue that by Corollary~\ref{cor1} since the original nonlinear model was structurally identifiable, the introduction of this new known forcing input function to scale $\beta_W$ did not change the structural identifiability.
DAISY output confirmed that \cref{siwr2} is still structurally identifiable as expected.

\section{Conclusion} \label{s:conclusions}

\noindent In this study, we have explored how introducing a known forcing input function into a system can improve its structural identifiability, and have proved that introducing a known forcing function input into a system will never worsen structural identifiability.
In proving results where the forcing function improves identifiabiltiy, this current work is limited to improving structural identifiability for a restricted subset of parameter combinations, where the coefficient map from the input-output relation contains coefficients made of additive or multiplicative parameter combinations.
More complex parameter combinations have yet to be addressed but would be a natural topic for future work, potentially leveraging numerical algebraic geometry approaches for handling larger systems \cite{bates2019identifiability}.

While for \citet{eisenberg2013examining} a forcing input function was successfully incorporated to improve practical identifiability, this kind of result is not guaranteed by the theory presented here, as we only address structural identifiability.
For example, analysis reveals that if the introduced data stream contains bias or excess error, the practical identifiability issue will not be resolved \cite{murphy2024measurement}. Further research into how noise and measurement error in forcing functions impact practical identifiability would be a valuable extension.

Additionally, our theoretical results apply specifically to ODE systems with rational function structure (polynomial quotients). Extension to systems with non-rational nonlinearities (exponential, trigonometric, logarithmic functions, etc.) remains an open question. Approximation-based methods (e.g., Padé or Taylor series) may enable extension to such systems, but theoretical guarantees would require additional analysis. This represents an important direction for future research.
%}

Despite these limitations, this methodology can already be applied to current problems in the literature, as demonstrated with the pharmacokinetics model in Section~\ref{ss:revisit}.
By utilizing algebraic solving software such as DAISY \cite{bellu2007daisy}, SIAN \cite{hong2019sian}, StructuralIdentifiability \cite{dong2023differential}, or STRIKE-GOLDD \cite{diaz2023strike}, coupled with these theorems, researchers can decide which parameters or subset of parameters need to be targeted for improved identifiability \textit{before} proceeding to the parameter estimation and forecasting steps.
This will reveal which parameters require more information, are more cost-effective to explore, or alternatively if a different model approach should be used instead.
Additionally, while most of our examples were drawn from biology/health applications, the applications of this theory extend beyond the biological, medical, and public health examples shown here.
This approach can provide researchers with a methodology to guarantee the structural identifiability and maintain the interpretability of their chosen model, thereby reinforcing the ethical imperative for robust and reliable scientific predictions that inform critical decision-making in public health, medicine, and beyond.

\section{Acknowledgements}
\noindent This work is the result of the research funded by the National Science Foundation  (grant no. DMS-1853032), the Clare Boothe Luce Program for Women in STEM, the Michigan Institute for Computational Discovery and Engineering, and Los Alamos National Laboratory Centers. The authors used AI assistance (Claude, Anthropic) for LaTeX formatting and document preparation. All mathematical results, proofs, theorems, biological interpretations, and scientific conclusions are entirely the work of the authors. AI was used solely for technical manuscript preparation and did not contribute to the research, analysis, or scientific content.

\printcredits

\bibliographystyle{cas-model2-names}

\bibliography{main}

\appendix
\section{Appendix}

\subsection{Characteristic Set Using Ritt's Pseudo-Division Example} \label{RPDexample}

\noindent This example illustrates a case of local but not global identifiability and allows us to demonstrate the characteristic set algorithm in more detail. 

\subsubsection{Original System}
\begin{figure}
    \centering
    \includegraphics[width=0.25\textwidth]{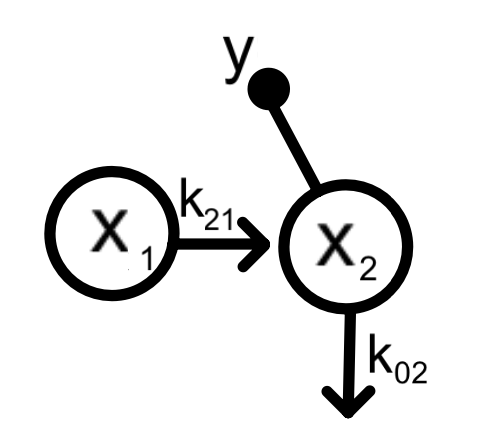}
    \caption{A 2-compartment model with a single rate flow from $x_1$ to $x_2$ and a leak out of $x_2$.}
    \label{fig:simple_appendix}
\end{figure}

\noindent Consider the following simple system of equations:
\begin{subequations}
 \begin{align} 
     \dot{x}_1 & = -k_{21} x_1 \label{exM1a_appendix} \\ 
     \dot{x}_2 & = k_{21} x_1 - k_{02} x_2; \label{exM1b_appendix} \\ 
     y & = x_2 \label{exM1c_appendix} 
 \end{align}
\end{subequations}
where our set of parameters is $\pmb{\theta} = \{ k_{21}, k_{02} \}$, $y$ is the measured output, and $x_1$ and $x_2$ are the state variables. 
We can visualize this model as the simple 2-compartment system in \Cref{fig:simple_appendix}.
Using the ranking system defined by \Cref{ranking}, we can see that for this system, the variables are ranked as: $$ y < x_1 < x_2 < \dot{x}_1 < \dot{x}_2 $$

To understand the characteristic set algorithm in detail, we walk through the steps of the algorithm for this example.

\noindent \textbf{0. Rewrite:} We first rewrite \Cref{exM2a_appendix}-\Cref{exM2c_appendix} such that all equations equal zero and order them by rank:
\begin{subequations}
 \begin{align} 
     0 & = y - x_2 & \text{leader: } x_2 \label{exM2a_appendix} \\
     0 & = \dot{x}_1 + k_{21} x_1  & \text{leader: } \dot{x}_1 \label{exM2b_appendix} \\ 
     0 & = \mathcolorbox{PowderBlue}{\dot{x}_2 - k_{21} x_1 + k_{02} x_2} & \text{leader: } \dot{x}_2 \label{exM2c_appendix} 
 \end{align}
\end{subequations}

For this system $n=2$ and $r=1$, for a total of 3 polynomials, one of which will eventually become our input-output relation in the final characteristic set.
Since \Cref{exM2c_appendix} contains the derivative of the leader of \Cref{exM2a_appendix}, we will begin the algorithm by comparing these.

\noindent \textbf{1. Compare:} We will consider \Cref{exM2a_appendix} as $A_j$ with the leader $x_2$, and compare it to \Cref{exM2c_appendix} which will be $A_i$, with leader $\dot{x}_2$, a derivative of the leader of $A_j$.

\noindent \textbf{2. Reduce:} We want to reduce $A_i$ with respect to $A_j$. 
$A_i$ contains the first derivative of $x_2$. 
Then we need to differentiate $A_j$ once to get $A_j^{(1)}:~~0 = \dot{y} - \dot{x}_2$. Divide $A_i$ by $A_j^{(1)}$ with respect to $\dot{x}_2$ to find pseudo-remainder $R$. 

\noindent \textbf{3. Replace:} Replace $A_i$ with $R$ and reorder the polynomials by rank. 
\begin{subequations}
 \begin{align} 
     0 & =  y - x_2 & \text{leader: } {x}_2 \label{exM3a_appendix} \\
     0 & =  \mathcolorbox{PowderBlue}{\dot{y} - k_{21} x_1 + k_{02} x_2} & \text{leader: } {x}_2 \label{exM3b_appendix} \\
     0 & =  \dot{x}_1 + k_{21} x_1 & \text{leader: } \dot{x}_1    \label{exM3c_appendix} 
 \end{align}
\end{subequations}
Note: When reviewing the algebraic rank of the new $A_i$ in \Cref{exM3b_appendix} versus $A_j$ in \Cref{exM3a_appendix}, we find that it is still of equal or higher rank. 

\noindent \textbf{2. Reduce (x2):} Divide $A_i: 0 = \dot{y} - k_{21} x_1 + k_{02} x_2 $ by $A_j: 0 = x_2 - y$ with respect to leader $x_2$ and find the new pseudo-remainder $R$.

\noindent \textbf{3. Replace (x2):} Replace $A_i$ with $R$ and reorder the polynomials by rank. 
As such, we have the new reduced system:
\begin{subequations}
 \begin{align} 
     0 & =  \mathcolorbox{PowderBlue}{\dot{y} + k_{02} y - k_{21} x_1} & \text{leader: } {x}_1 \label{exM4a_appendix} \\ 
     0 & =  y - x_2 & \text{leader: } {x}_2 \label{exM4b_appendix} \\
     0 & =   \mathcolorbox{MistyRose}{\dot{x}_1 + k_{21} x_1} & \text{leader: } \dot{x}_1 \label{exM4c_appendix} 
 \end{align}
\end{subequations}

\noindent \textbf{1. Compare:} Now we can see that \Cref{exM4a_appendix} is reduced with respect to \Cref{exM4b_appendix}. 
\Cref{exM4c_appendix} however contains the derivative of the leader of \Cref{exM4a_appendix}. Let \Cref{exM4c_appendix} be $A_i$ and \Cref{exM4a_appendix} be $A_j$.

\noindent \textbf{2. Reduce:} $A_i$ contains the first derivative of $x_1$. 
We differentiate $A_j$ once to obtain $A_j^{(1)}:~~ 0 = \ddot{y} - k_{21} \dot{x}_1 + k_{02} \dot{y}$.
Note that $\dot{x}_1$ has a coefficient in $A_j^{(1)}$, so we need to multiply $A_i$ by $-k_{21}$ before proceeding.
So we will divide $-k_{21} A_i$ by $A_j^{(1)}$ with respect to $\dot{x}_1$, and keep the pseudo-remainder $R$.

\noindent \textbf{3. Replace:} We want to replace $A_i$ with $R$.
\begin{subequations}
\begin{align} 
    0 & =  \dot{y} + k_{02} y - k_{21} x_1& \text{leader: } {x}_1 \label{exM5a_appendix} \\
    0 & = \mathcolorbox{MistyRose}{\ddot{y} + k_{02}\dot{y} + k_{21}^2 x_1} & \text{leader: } x_1 \label{exM5b_appendix} \\ 
    0 & =  x_2 - y & \text{leader: } {x}_2 \label{exM5c_appendix} 
\end{align}
\end{subequations}
Note: \Cref{exM5b_appendix} is still not reduced with respect to \Cref{exM5a_appendix} when we review the algebraic ranking. Let \Cref{exM5b_appendix} be $A_i$ and \Cref{exM5a_appendix} be $A_j$.

\noindent \textbf{2. Reduce (x2):} Divide $A_i: 0 = \ddot{y} + k_{02}\dot{y} + k_{21}^2 x_1$ by $A_j: \dot{y} + k_{02} y - k_{21} x_1$ with respect to $x_1$ to find the pseudo-remainder $R$. 

\noindent \textbf{3. Replace (x2):} We now replace $A_i$ with $R$ and reorder the polynomials by their ranking:
\begin{subequations}
\begin{align} 
    0 & = \mathcolorbox{MistyRose}{\frac{1}{k_{21}}\ddot{y} + (\frac{k_{02}}{k_{21}} + 1)\dot{y} + k_{02} y} & \text{leader: } \ddot{y} \label{exM6a_appendix} \\ 
    0 & =  \dot{y} - k_{21} x_1 + k_{02} y & \text{leader: } {x}_1 \label{exM6b_appendix} \\ 
    0 & =  y - x_2 & \text{leader: } {x}_2 \label{exM6c_appendix} 
\end{align}
\end{subequations}

\vspace{0.2cm}
Having reduced all state variables, we now normalize the input-output relation in the final step.

\noindent \textbf{4. Enforce Monic $\pmb{A_i}$:} We can see that the leader of the lowest ranked polynomial (\Cref{exM6a_appendix}) has a coefficient $1/k_{21}$, so we will divide the polynomial by this leading coefficient.
Thus, the final decomposition of the system (\ref{exM1a_appendix})- (\ref{exM1c_appendix}) is:
\begin{subequations}
\begin{align} 
    0 & = \mathcolorbox{MistyRose}{\ddot{y} + (k_{02} + k_{21})\dot{y} + k_{02}k_{21} y} & \text{leader: } \ddot{y} \label{io1a_appendix} \\ 
    0 & =  \dot{y} - k_{21} x_1 + k_{02} y & \text{leader: } {x}_1 \label{io1b_appendix}\\ 
    0 & =  y - x_2 & \text{leader: } {x}_2 \label{io1c_appendix}
\end{align}
\end{subequations}
As the lowest ranked polynomial \Cref{io1a_appendix} no longer contains any state variables $\pmb{x}$, it is our \textit{input-output relation} for system (\ref{exM1a_appendix})- (\ref{exM1c_appendix}).
Taking the coefficients of this equation, we generate the coefficient map $c(\pmb{\theta}) = \{k_{02} + k_{21}, k_{02}k_{21} \}$, which provides two unique parameter set solutions for system (\ref{exM1a_appendix})- (\ref{exM1c_appendix}).
The work can be shown as the following:
\begin{align*}
    k_{02} + k_{21} & = a_1 \\
    k_{02}k_{21} & = a_2 \\
    \Longrightarrow \\
    k_{21} & = k_{02} - a_1\\
    k_{02} & = \frac{a_1 \pm \sqrt{a_1^2 - 4a_2}}{2}
\end{align*}

\subsubsection{Augmented System with Forcing Function}\label{asec:2compex}
\noindent If the user wanted to find a single unique parameter set solution to this system, 
usually constraints on the parameter domain reveal that only one of the two solutions is realistic.
However, we illustrate in this example how including a scaling forcing input function $u(t)$ in the system both (a) does not worsen the system's structural identifiability and (b) can in some cases improve it. 
In this exercise, we introduce the forcing input function $u(t)$ that satisfies the definition above to scale the parameter $k_{21}$.
This condition guarantees nonzero derivatives of $u(t)$ as it is smooth and non-constant.

Then our new system of equations is:
\begin{subequations}
 \begin{align} 
     \dot{x}_1 & = -k_{21} \mathcolorbox{Yellow}{u} x_1 \label{eq1a_appendix}\\
     \dot{x}_2 & = k_{21} \mathcolorbox{Yellow}{u} x_1 - k_{02} x_2; \label{eq1b_appendix}\\ 
     y & = x_2 \label{eq1c_appendix}
 \end{align}
\end{subequations}
The overall rankings for this system are: $$ u < y < x_1 < x_2 < \dot{x}_1 < \dot{x}_2 $$

\noindent \textbf{0. Rewrite:} We first rewrite \Cref{eq1a_appendix}-\Cref{eq1b_appendix} such that all equations equal zero and order them by rank:
\begin{subequations}
 \begin{align} 
     0 & = x_2 - y& \text{leader: } x_2 \label{eq2a_appendix}\\
     0 & = \dot{x}_1 + k_{21} \mathcolorbox{Yellow}{u} x_1  & \text{leader: } \dot{x}_1 \label{eq2b_appendix}\\ 
     0 & = \mathcolorbox{PowderBlue}{\dot{x}_2 - k_{21} \mathcolorbox{Yellow}{u} x_1 + k_{02} x_2} & \text{leader: } \dot{x}_2 \label{eq2c_appendix}
 \end{align}
\end{subequations}

For this system $n=2$ and $r=1$, for a total of 3 polynomials, one of which will be our input-output relation in the final characteristic set.
Since \Cref{eq2c_appendix} contains the derivative of the leader of \Cref{eq2a_appendix}, we will begin the algorithm by comparing these.

\noindent \textbf{1. Compare:} We will consider \Cref{eq2a_appendix} as $\tilde{A}_j$ with the leader $x_2$, and compare it to \Cref{eq2c_appendix} which will be $\tilde{A}_i$, with leader $\dot{x}_2$, a derivative of the leader of $\tilde{A}_j$.

\noindent \textbf{2. Reduce:} We want to reduce $\tilde{A}_i$ with respect to $\tilde{A}_j$. 
$\tilde{A}_i$ contains the first derivative of $x_2$. 
Then we need to differentiate $\tilde{A}_j$ once to get $\tilde{A}_j^{(1)}:~~0 = \dot{y} - \dot{x}_2$. Divide $\tilde{A}_i$ by $\tilde{A}_j^{(1)}$ with respect to $\dot{x}_2$ to find pseudo-remainder $R$. 

\noindent \textbf{3. Replace:} Replace $\tilde{A}_i$ with $R$ and reorder the polynomials by rank. 
\begin{subequations}
 \begin{align}
     0 & =  x_2 - y & \text{leader: } {x}_2 \label{eq3a_appendix} \\
     0 & =  \mathcolorbox{PowderBlue}{\dot{y} - k_{21} \mathcolorbox{Yellow}{u} x_1 + k_{02} x_2} & \text{leader: } {x}_2 \label{eq3b_appendix} \\
     0 & =  \dot{x}_1 + k_{21} \mathcolorbox{Yellow}{u} x_1 & \text{leader: } \dot{x}_1  \label{eq3c_appendix}   
 \end{align}
\end{subequations}
Note: When reviewing the algebraic rank of the new $\tilde{A}_i$ in \Cref{eq3b_appendix} versus $\tilde{A}_j$ in \Cref{eq3a_appendix}, we find that it is still higher rank. 

\noindent \textbf{2. Reduce (x2):} Divide $\tilde{A}_i: 0 = \dot{y} - k_{21} u x_1 + k_{02} x_2 $ by $\tilde{A}_j: 0 = x_2 - y$ with respect to leader $x_2$ and find the new pseudo-remainder $R$.

\noindent \textbf{3. Replace (x2):} Replace $\tilde{A}_i$ with $R$ and reorder the polynomials by rank. 
As such, we have the new reduced system:
\begin{subequations}
 \begin{align} 
     0 & =  \mathcolorbox{PowderBlue}{\dot{y} - k_{21} \mathcolorbox{Yellow}{u} x_1 + k_{02} y} & \text{leader: } {x}_1 \label{eq4a_appendix} \\ 
     0 & =  x_2 - y & \text{leader: } {x}_2 \label{eq4b_appendix} \\
     0 & =   \mathcolorbox{MistyRose}{\dot{x}_1 + k_{21} \mathcolorbox{Yellow}{u} x_1} & \text{leader: } \dot{x}_1 \label{eq4c_appendix} 
 \end{align}
\end{subequations}

\noindent \textbf{1. Compare:} Now we can see that \Cref{eq4a_appendix} is reduced with respect to \Cref{eq4b_appendix}. 
\Cref{eq4c_appendix} however contains the derivative of the leader of \Cref{eq4a_appendix}, $\dot{x}_1$. Let \Cref{eq4c_appendix} be $\tilde{A}_i$ and \Cref{eq4a_appendix} be $\tilde{A}_j$.

\noindent \textbf{2. Reduce:} 
$\tilde{A}_i$ contains the first derivative of $x_1$. 
We differentiate $\tilde{A}_j$ once to obtain $\tilde{A}_j^{(1)}:~~ 0 = \ddot{y} + k_{02} \dot{y} - k_{21} \dot{u} {x}_1 - k_{21} u \dot{x}_1$.
Now we have gained an extra term that did not exist in the original steps.
And as before, we need to multiply $\tilde{A}_i$ by the leading coefficient of $\tilde{A}_j^{(1)}$ (which is $-k_{21}$) before proceeding.
So we will divide $-k_{21} \tilde{A}_i$ by $\tilde{A}_j^{(1)}$ with respect to $\dot{x}_1$, and keep the pseudo-remainder $R$.

\noindent \textbf{3. Replace:} We want to replace $\tilde{A}_i$ with $R$.
\begin{fleqn}
\begin{subequations}
\begin{align} 
    0 & =  \dot{y} - k_{21} \mathcolorbox{Yellow}{u} x_1 + k_{02} y & \text{leader: } {x}_1 \label{eq5a_appendix}  \\
    0 & = \mathcolorbox{MistyRose}{\ddot{y} + k_{02}\dot{y}  - k_{21} \mathcolorbox{Yellow}{\dot{u}} x_1- k_{21}^2 \mathcolorbox{Yellow}{u^2} x_1} & \text{leader: } x_1 \label{eq5b_appendix}\\ 
    0 & =  x_2 - y & \text{leader: } {x}_2  \label{eq5c_appendix}
\end{align}
\end{subequations}
\end{fleqn}
Note: \Cref{eq5b_appendix} is still not reduced with respect to \Cref{eq5a_appendix} when we review the algebraic ranking. Let \Cref{eq5b_appendix} be $\tilde{A}_i$ and \Cref{eq5a_appendix} be $\tilde{A}_j$. Two terms in $\tilde{A}_i$ contain the leader, $x_1$, but the term containing $\dot{u}$ is higher ranked so we will reduce with respect to this term first.

\noindent \textbf{2. Reduce (x2):} We want to divide $\tilde{A}_i$ by $\tilde{A}_j$ with respect to $x_1$. 
The leader of $\tilde{A}_j$ in $x_1$ has coefficient $-k_{21}$, and so does the leader of $\tilde{A}_i$, but there is an added $u$ term that we need to multiply by before proceeding to find the pseudo-remainder $R$.
After reduction, we get:
\begin{subequations}
\begin{align} 
    0 & =  \dot{y} - k_{21} \mathcolorbox{Yellow}{u} x_1 + k_{02} y & \text{leader: } {x}_1 \label{eq6a_appendix}  \\
    0 & = \mathcolorbox{MistyRose}{\mathcolorbox{Yellow}{u}\ddot{y} + k_{02}\mathcolorbox{Yellow}{u}\dot{y} - \mathcolorbox{Yellow}{\dot{u}} \dot{y}  } & \text{leader: } x_1 \label{eq6b_appendix}\\
    &~~~~~\mathcolorbox{MistyRose}{- k_{02} \mathcolorbox{Yellow}{\dot{u}} y - k_{21}^2 \mathcolorbox{Yellow}{u^3} x_1} &  \nonumber \\ 
    0 & =  x_2 - y & \text{leader: } {x}_2  \label{eq6c_appendix}
\end{align}
\end{subequations}
Note: \Cref{eq6b_appendix} is \textit{still} not reduced with respect to \Cref{eq6a_appendix} when we review the algebraic ranking. Let \Cref{eq6b_appendix} be $\tilde{A}_i$ and \Cref{eq6a_appendix} be $\tilde{A}_j$.\red

\noindent \textbf{2. Reduce (x3):} We want to divide $\tilde{A}_i$ by $\tilde{A}_j$ with respect to $x_1$. 
Since the leader of $\tilde{A}_j$ in $x_1$ has coefficient $k_{21}$, we multiply $\tilde{A}_i$ by $1/k_{21}$ before proceeding to find the pseudo-remainder $R$.
After reduction, we get:
\begin{fleqn}
\begin{subequations}
\begin{align} 
    0 & = \mathcolorbox{MistyRose}{\frac{1}{k_{21}}\mathcolorbox{Yellow}{u}\ddot{y} + \frac{k_{02}}{k_{21}} \mathcolorbox{Yellow}{u}\dot{y} - \frac{1}{k_{21}}\mathcolorbox{Yellow}{\dot{u}} \dot{y}  } & \text{leader: } \ddot{y} \label{io2a_appendix_corrected} \\ 
    & ~~~~~\mathcolorbox{MistyRose}{ - \mathcolorbox{Yellow}{u^2} \dot{y} - \frac{k_{02}}{k_{21}} \mathcolorbox{Yellow}{\dot{u}} y + k_{02} \mathcolorbox{Yellow}{u^2}  y  } &  \nonumber \\ 
    0 & =  \dot{y} - k_{21} \mathcolorbox{Yellow}{u} x_1 + k_{02} y & \text{leader: } {x}_1 \label{io2b_appendix}\\ 
    0 & =  y - x_2 & \text{leader: } {x}_2 \label{io2c_appendix}
\end{align}
\end{subequations}
\end{fleqn}

\vspace{0.2cm}
With all state variables eliminated from the lowest-ranked polynomial, we complete the derivation by enforcing the monic condition.

\noindent \textbf{4. Enforce Monic $\pmb{\tilde{A}_i}$:} We divide by the coefficient of $\ddot{y}$ in \cref{io2a_appendix_corrected}.
The input-output relation derived via DAISY for system (\ref{eq1a_appendix})-(\ref{eq1c_appendix}) is:
\begin{fleqn}
\begin{subequations}
\begin{align} 
    0 & = \mathcolorbox{MistyRose}{\mathcolorbox{Yellow}{u}\ddot{y} + k_{02} \mathcolorbox{Yellow}{u}\dot{y} + k_{21} \mathcolorbox{Yellow}{u^2} \dot{y} - \mathcolorbox{Yellow}{\dot{u}}\dot{y} } & \text{leader: } \ddot{y} \label{io2a_appendix_daisy}\\ 
    & ~~~~~\mathcolorbox{MistyRose}{ - k_{02}\mathcolorbox{Yellow}{\dot{u}} y + k_{02}k_{21}\mathcolorbox{Yellow}{u^2} y } &  \nonumber \\ 
    0 & =  \dot{y} - k_{21} \mathcolorbox{Yellow}{u} x_1 + k_{02} y & \text{leader: } {x}_1 \label{io2b_daisy}\\ 
    0 & =  y - x_2 & \text{leader: } {x}_2 \label{io2c_daisy}
\end{align}
\end{subequations}
\end{fleqn}

%\textcolor{green}{$$ u\ddot{y} -\dot{u}\dot{y}  + k_{21} u^2 \dot{y} + k_{02}u \dot{y} - k_{02}\dot{u}y + k_{02}k_{21}u^2 y$$}

Therefore our new coefficient map is $\tilde{c}(\pmb{\theta}) = \{ k_{02}, k_{21}, \\k_{02}k_{21}\}$.
From this, the coefficients $k_{02}$ and $k_{21}$ are identifiable (since $u$ is known and non-zero).
Thus, there is only one unique parameter set solution $\{k_{02}, k_{21}\}$.
As such, the introduction of a forcing input function $u$ allowed us to gain information about the original system (\ref{exM1a_appendix})- (\ref{exM1c_appendix}) and achieve global identifiability.

If for the sake of argument we assume $u$ was a constant function ($u=C, \dot{u}=0$), then the input-output relation of \Cref{io2a_appendix_daisy} reduces back to the original input-output relation of \Cref{io1a_appendix} after enforcing monic $\tilde{A}_i$. 
This illustrates that no existing information in the system was lost due to introduction of $u$.
That is, the parameter solution set from \Cref{io1a_appendix} is contained within that of \Cref{io2a_appendix_daisy}.

\setlength{\leftskip}{0.5cm}
\vspace{0.2cm}
\noindent \textit{Remark A1. }
    Given a coefficient map $c(\pmb{\theta})$, we refer to this type of parameter subset $\{\theta_i + \theta_j, \theta_i \theta_j \} \subset c(\pmb{\theta})$ as a \textit{sum-product relationship} between $\theta_i$ and $\theta_j$, where one or two solutions are expected for this parameter subset.
 \setlength{\leftskip}{0cm}

\subsection{Forcing functions can improve identifiability: linear and bilinear compartmental models (Appendix)} \label{s:3dim_appendix}

\noindent In this section we specifically consider the case of small, linear or bilinear compartmental models where all rates between compartments are constant with only a single observed state, and show when the models are structurally identifiable, and when introducing a forcing input function will improve the structural identifiability from a mathematical perspective.

We note that  the examples in this section are not solved via a general proof, rather all possible forms of each model were run exhaustively with an automated shell script using DAISY. For each size model (1, 2, and 3 compartments), the full models with all terms was first run and then parameters were systematically removed to form each potential sub-case, and the input-output relation was regenerated.
Here we provide a general summary of the findings for each case, with sample code provided in Appendix \ref{DAISYex} to run any individual model as needed.

\subsubsection{Linear Models up to 3D} \label{ss:lin_appendix}

\noindent For the linear model, we take advantage of the $\dot{\pmb{x}} = A\pmb{x} + \pmb{b}$ structure, initially generating a model with movement into/out of all compartments by all compartments.
To constrain and simplify the model, we impose $\pmb{b} = 0$, such that there are no existing constant inputs into the system.
This starting system is assumed to have no existing time-varying input functions.
Finally, we constrain the measured component to only one state variable, of the form $y = x_1$.

Examples of linear models can be found in the literature, beginning with the pharmacokinetics model that we showed in Section~\ref{ss:MotEx2_diff_alg}. 
Other examples include but are not limited to damped harmonic oscillators, radiation decay chains, or circuit design \cite{belyshev2014studying, bauer2001transport,chandrasekar2007lagrangian, dekker1981classical, fu2021solving, bellen1999methods, lohfink2005linear}.

Here we explore all possible structural identifiabilities for linear systems up to dimension 3.
Let $n: n \in \{1,2,3\}$ be the dimension of the linear system, and $\pmb{\theta}$ represent the arbitrary real-valued parameter vector, not dependent on time.

\vspace{0.2cm}
\noindent \textbf{Case 1.} If $n=1$, then the system contains only one state equation and at most can uniquely identify 1 parameter without initial conditions.
We can generalize the formula of this system to be: $\dot{x} = \theta_1 x$.
Since there is only one state variable to measure, we can generalize the measurement function to be $y = x$.
Then the only input-output relation this system can have is:
\begin{align*}
    0 & = \dot{y} - \theta_1 y
\end{align*}
This system is already structurally identifiable and does not need further inputs to be incorporated.

\vspace{0.2cm}
\noindent \textbf{Case 2.} If $n=2$, the system now contains two state equations and can uniquely identify up to 2 parameters without knowledge of the initial conditions.
We generalize this system as:
\begin{align} \label{eq:2d_a_appendix}
    \dot{x}_1 &= \theta_1 x_1 + \theta_2 x_2 \\ \nonumber
    \dot{x}_2 &= \theta_3 x_1 + \theta_4 x_2 \\ \nonumber
    y &= x_1
\end{align}
assuming that only one state at most is measurable.
Then the input-output relation will be:
\begin{align}\label{eq:io_2d_appendix}
        0 = \ddot{y} - (\theta_1 + \theta_4) \dot{y} + (\theta_1 \theta_4 - \theta_2 \theta_3) y
\end{align}

%$$df(y,t,2) - df(y,t)*(p1 + p4) + y*(p1*p4 - p2*p3)$$

The case for $x_2$ being the observed state yields the same input-output relation (due to symmetry).
Further, in the case where $\theta_{1,4}$ and $\theta_{2,3}$ represent out flows and in flows respectively, the coefficient map for the input-output relation remains the same.
This relation is unidentifiable if all parameters are distinct and nonzero.

While the identifiable combinations of this system cannot be immediately resolved to yield global identifiability, we can do so with two alternative applications of the methods presented here (scaling or replacing two parameters with inputs), as illustrated in Figure~\ref{fig:LinearCase2_generic} (in Section~\ref{s:generic_ex_main}).

Next we consider cases where various parameters are removed from the system. After re-calculating the input-output equation for each case where one or more parameters were removed, we found broadly similar patterns in the unidentifiability of the system and approaches to resolve identifiability by forcing function inputs. We illustrate two examples:
\begin{itemize} 
    \item \textit{If $\theta_2$ or $\theta_3$ removed:} then up to two unique solutions for $\{\theta_1, \theta_4\}$ are possible, but the model is still unidentifiable since $\theta_2$ or $\theta_3$ do not appear in the input-output relation. In this case, scaling nor replacing either $\theta_2$ or $\theta_3$ with a forcing input function will \textit{not} resolve the structural identifiability issue.
    
    \item \textit{If $\theta_1$ and $\theta_3$ removed (or $\theta_2$ and $\theta_4$ removed):} then up to two unique solutions for $\{\theta_2, \theta_4\}$ (or $\{\theta_1, \theta_3\}$) are possible, and the model is structurally identifiable. Scaling either remaining parameter with a forcing input function will generate a single unique parameter set, and thus make the model globally structurally identifiable with known initial conditions. 
    
    \noindent \textit{Remark A2. } 
        Replacing $\theta_2$ or $\theta_3$ with a known forcing input function will also produce the same effect as removing $\theta_2$ or $\theta_3$.
\end{itemize}
In this way, for a 2 dimensional linear compartment model with constant rate flows and no existing inputs, we have summarized all cases where scaling or replacing a parameter with a known forcing input function can possibly improve the structural identifiability of the model.

\vspace{0.2cm}
\noindent \textbf{Case 3.} If $n=3$, the system now contains three state equations and can uniquely identify up to 3 parameters without initial conditions. 
We generalize this system as:
\begin{align} \label{eq:3d_appendix}
    \dot{x}_1 &= \theta_1 x_1 + \theta_2 x_2  + \theta_3 x_3 \\ \nonumber
    \dot{x}_2 &= \theta_4 x_1 + \theta_5 x_2 + \theta_6 x_3\\ \nonumber
    \dot{x}_3 &= \theta_7 x_1 + \theta_8 x_2 + \theta_9 x_3 \\ \nonumber
    y &= x_1
\end{align}
assuming that only one state at most is measured.
The input-output relation for this generalized system is:
\begin{fleqn}
\begin{multline} \label{eq:io_3d_appendix}
    - \dddot{y} + (\theta_1 + \theta_5 + \theta_9) \ddot{y} \\
    + ( - \theta_1 \theta_5 - \theta_1 \theta_9 + \theta_2 \theta_4 + \theta_3 \theta_7 - \theta_5 \theta_9 + \theta_6 \theta_8) \dot{y} \\
    + \big[\theta_1 (\theta_5 \theta_9 - \theta_6 \theta_8) - \theta_2 (\theta_4 \theta_9 - \theta_6 \theta_7) \\ + \theta_3 (\theta_4 \theta_8 - \theta_5 \theta_7) \big] y = 0
\end{multline}
\end{fleqn}

The case for $x_2$ or $x_3$ being the partially observed state was confirmed in DAISY to yield the same input-output relation (and is clear due to symmetry).

First, we can observe the following terms within the input-output coefficients (not as identifiable combinations themselves, but as components of them): $\{\theta_1$, $\theta_5 + \theta_9$, $\theta_5 \theta_9$, $\theta_2 \theta_4$, $\theta_3 \theta_7$ , $\theta_6 \theta_8$,
$\theta_2 \theta_6 \theta_7$ , $\theta_3 \theta_4 \theta_8\}$. 
Currently, all of the parameters are unidentifiable. The term combinations shown here are just examples designed to emphasize what parameter combinations can be targeted for scaling and replacement with known forcing function or constant inputs.

For example, we see that $\theta_1 + \theta_5 + \theta_9$ form an additive combination, which from Theorem \ref{scaling_1d} we could potentially resolve with two applications of input scaling.

If initial conditions are known, $\theta_3$ and $\theta_7$ also become identifiable.

Therefore, from this original system it may be possible to identify a finite set of solutions for the parameter subset $\{ \theta_1, \theta_3, \theta_5, \theta_7, \theta_9\}$, but this still leaves half of the original parameter set unaccounted for.

\begin{figure}
    \centering
    \includegraphics[width=0.12\textwidth]{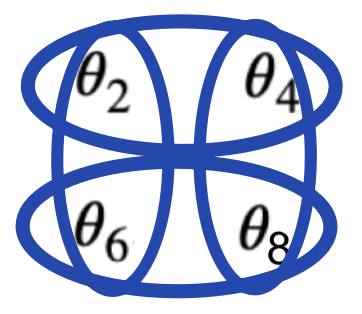}
    \caption{Parameter graph of unidentifiable parameters for the 3D Linear model when all parameters are nonzero. All parameters shown have a functional relationship with at least two other parameters, and cannot be separated as they are currently. }
    \label{fig:3d_orig_paramgraph_appendix}
\end{figure}

The relationships between the remaining unidentifiable parameters from this set $\{\theta_2, \theta_4, \theta_6, \theta_8\}$ can be visualized in \cref{fig:3d_orig_paramgraph_appendix}.

These remaining parameters form a connected component of multiplicative relationships, visualized in \cref{fig:3d_orig_paramgraph_appendix}.
\textit{Replacement} of all these parameters in the subset $\{\theta_2, \theta_4, \theta_6, \theta_8 \}$ will make our system structurally identifiable by Theorem~\ref{replace_1d_FINAL}.
\textit{Scaling} is not viable for this system to resolve the unidentifiability, as these parameters are in a multiplicative relationship that violates the conditions to use Theorem~\ref{scaling_1d}.

\vspace{0.2cm}
\noindent Therefore, for up to 3 dimensions we have demonstrated with generality when linear models with constant rate flows can improve their structural identifiability through replacing or scaling a parameter with a known forcing input function.

\subsubsection{Bilinear Mixing Models up to 2D} \label{ss:bilin_appendix}
\noindent We can begin to see from the linear model discussion that even with the constraints of constant rate flows, a single observed variable, and no existing input functions, this problem does not scale well.
Even so, we begin the discussion for what we will refer to as \textit{bilinear models}.

A bilinear model is a model of the form $\dot{\pmb{x}} = A f(\pmb{x}) + \pmb{b}$, where $\pmb{b}$ is a constant forcing input, $\pmb{x} \in \mathbb{R}^n$ contains all state variables, and $f(\pmb{x})$ is a vector of all the state variables \textit{and} their combinations of the form $x_i x_j$ given $i,j \in \{1, ...,n\}$.
We will focus on a subset of this model class that only includes mixing terms, such that we impose the constraint $x_i x_j$ given $i \neq j$.
To further constrain our model, we set $b = 0$ so the systems considered do not contain any existing input function information.
As with \cref{ss:lin_appendix}, we assume only a single state variable is observed, such that $y = \theta_{m} x_i$ where $m$ is the total number of parameters in the system.

Examples of models with bilinear mixing can be found in the literature, from the SIR models used in public health analysis, to Lotka-Volterra models for population monitoring, to the Lorenz system. 
An example variation of an SIR model is explored in more detail in Section~\ref{ss:cholera}.

\vspace{0.2cm}
\noindent \textbf{Case 1.}
If $n=1$, then the system contains one state variable, and is generalized as:
\begin{align*}
    \dot{x} &= \theta_1 x\\ \nonumber
    y & = \theta_2 x
\end{align*}
Note this is the same as Case 1 in \cref{ss:lin_appendix} as a single state variable cannot have mixing as defined here. However, we see that in calculating the input-output equation, the parameter $\theta_2$ does not and so is not identifiable without adding an input or information about the initial conditions.

\vspace{0.2cm}
\noindent \textbf{Case 2.}
If $n=2$, then the system contains two state variables and can be generalized as the following:
\begin{align} \label{bilin2d_appendix}
    \dot{x}_1 &= \theta_1 x_1 + \theta_2 x_1 x_2 + \theta_3 x_2 \\ \nonumber
    \dot{x}_2 &= \theta_4 x_1 + \theta_5 x_1 x_2 + \theta_6 x_2 \\ \nonumber
    y &= \theta_7 x_1
\end{align}
Then the input-output relation is:
\begin{fleqn}
\begin{multline} \label{eq:bilin_io_2d_appendix_full}
    \theta_2  \theta_7 \ddot{y} y + \theta_3  \theta_7^2 \ddot{y}  
    -\theta_2  \theta_7 \dot{y}^2 
    - \theta_2  \theta_5 \dot{y} y^2 \\
    - \theta_7 ( \theta_2  \theta_6 +  \theta_3  \theta_5) \dot{y} y 
    - \theta_3  \theta_7^2 ( \theta_1 +  \theta_6) \dot{y} \\
    + \theta_2 ( \theta_1  \theta_5 -  \theta_2  \theta_4) y^3 
    + \theta_7 ( \theta_1  \theta_2  \theta_6 +  \theta_1  \theta_3  \theta_5 - 2  \theta_2  \theta_3  \theta_4) y^2  \\+ 
    \theta_3  \theta_7^2 ( \theta_1  \theta_6 -  \theta_3  \theta_4) y  = 0
\end{multline}
\end{fleqn}

% \begin{comment}
% \begin{multline} 
% \ddot{y}y\theta_2\theta_7 + \ddot{y}\theta_3\theta_7^2 \\- \dot{y}^2\theta_2\theta_7 - \dot{y}y^2\theta_2\theta_5 \\- \dot{y}y\theta_7(\theta_2\theta_6\\
%  + \theta_3\theta_5) - \dot{y}\theta_3\theta_7^2(\theta_1 + \theta_6)\\ + y^3\theta_2(\theta_1\theta_5 - \theta_2\theta_4) \\+ y^2\theta_7(\theta_1\theta_2\theta_6 + \theta_1\theta_3\theta_5 - 2\theta_2\theta_3\theta_4) \\
% + y\theta_3\theta_7^2(\theta_1\theta_6 - \theta_3\theta_4) = 0
% \end{multline}
% \end{fleqn}
% \end{comment}

The relationships between the different parameters of this system are not immediately obvious.
With known initial conditions it can be shown that the above model is globally structurally identifiable as it is. 
Unlike the linear model 2D case, this system has only one unique parameter set given when given initial conditions, so no further scaling or replacement is required for unique structural identifiability if all parameters are nonzero.

\begin{figure}
  \centering
  \begin{tabular}[b]{c c c}
    \includegraphics[height = 1.6cm]{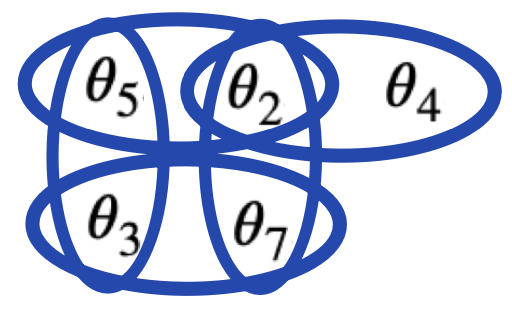} &
    \includegraphics[height = 1.6cm]{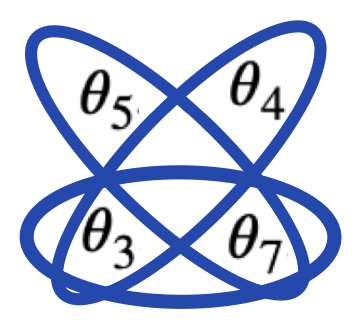} &
    \includegraphics[height = 1.6cm]{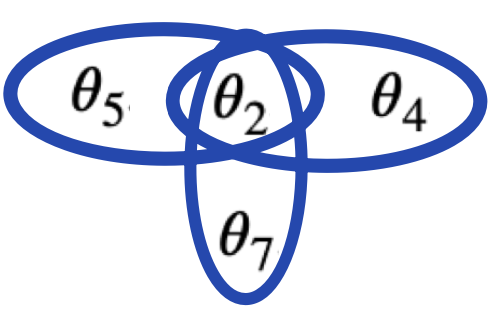}\\
    \small (a) $\theta_1$ removed & \small (b) $\theta_2 $ removed & \small (c) $\theta_3 $ removed\\
    \includegraphics[height = 1.6cm]{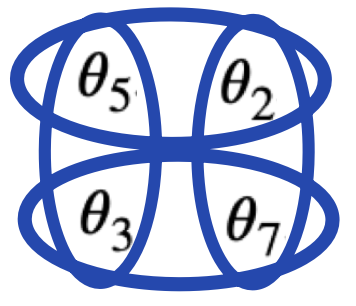} &
    \includegraphics[height = 1.6cm]{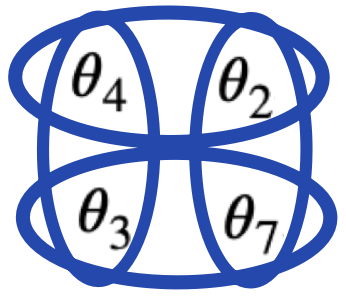} &
    \includegraphics[height = 1.6cm]{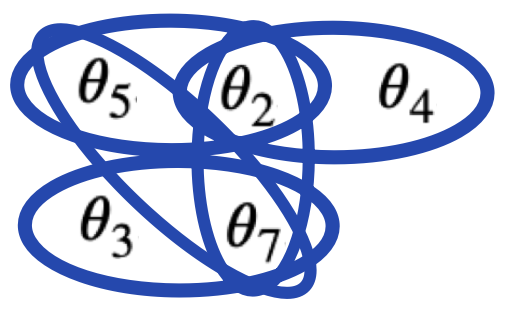} \\
    \small (d) $\theta_4 $ removed & \small (e) $\theta_5 $ removed& \small (f) $\theta_6 $ removed
  \end{tabular} 
  \caption{Comparison of parameter graphs generated by sub-cases for bilinear mixing 2D model. This allows researchers to determine visually which parameters to potentially target for replacement or scaling for improving identifiability.}
  \label{fig:bilin_paramgraph}
\end{figure}

Let us then consider sub-cases where one of the above parameters is removed from the original system (similar analysis can be used to explore cases where more than one parameter is removed, here we just illustrate a few examples). The input-output equations for these sub-cases are derived using DAISY.

We can visualize how replacing a parameter will change our parameter relationships by using the parameter graph. 
In \cref{fig:howto} (in Section~\ref{s:forcefunc}), we show step by step how replacing $\theta_4$ in the $\theta_2 $ removed sub-case (originally visualized in \cref{fig:bilin_paramgraph}(b)) can make the overall system structurally identifiable. 

\subsection{Example DAISY Input File and Terminal Output} \label{DAISYex}

\begin{lstlisting}[language={},
caption={DAISY input file for the 2-compartment model.}]
% State variables:
vars x1, x2;

% Parameters:
params k21, k02;

% Equations of the model:
odes
    x1' = -k21 * x1;
    x2' = k21 * x1 - k02 * x2;

% Output equation(s):
output y = x2;

% Solve for identifiability:
solve_globally; 
\end{lstlisting}

\begin{lstlisting}[language=bash, caption=Example terminal output from DAISY for the original model., basicstyle=\ttfamily\footnotesize, backgroundcolor=\color{black!5}, commentstyle=\color{black!60}]
DAISY (Version X.Y.Z)

Reading model from input file...
Model successfully parsed.
Number of state variables: 2
Number of parameters: 2
Number of output functions: 1

Starting identifiability analysis (global)...
Computing characteristic set...
Input-output equation(s):
y'' + (k02+k21)*y' + k02*k21*y = 0

Analyzing coefficients for identifiability...
The model is LOCALLY identifiable.
Identifiable combinations:
k02 + k21
k02 * k21

Parameters k21, k02 have 2 solution(s).

Global identifiability check:
The model is NOT GLOBALLY identifiable.

Identifiability analysis complete.
\end{lstlisting}

\begin{lstlisting}[language={}, caption={DAISY input file for the augmented 2-compartment model with input function u}]
% State variables:
vars x1, x2;

% Parameters:
params k21, k02;

% Known inputs (forcing functions):
inputs u; 

% Equations of the model:
odes
    x1' = -k21 * u * x1;
    x2' = k21 * u * x1 - k02 * x2;

% Output equation(s):
output y = x2;

% Solve for identifiability:
solve_globally;
\end{lstlisting}

\begin{lstlisting}[language=bash, caption=Example terminal output from DAISY for the augmented model., basicstyle=\ttfamily\footnotesize, backgroundcolor=\color{black!5}, commentstyle=\color{black!60}]
DAISY (Version X.Y.Z)

Reading model from input file...
Model successfully parsed.
Number of state variables: 2
Number of parameters: 2
Number of inputs: 1 (u)
Number of output functions: 1

Starting identifiability analysis (global)...
Computing characteristic set...
Input-output equation(s):
u*y'' + (k02*u + k21*u^2 - u')*y' 
+ (k02*k21*u^2 - k02*u')*y = 0 

Analyzing coefficients for identifiability...
The model is GLOBALLY identifiable.
Parameters k21, k02 are globally identifiable.

Identifiability analysis complete.
\end{lstlisting}

\vspace{1cm}

\subsection{Additional Lemma 1 Proof Information} \label{prop1proof}

\noindent \textbf{Lemma 1.} 
\textit{Let $\tilde{u}$ be a forcing input function, and let $M$ be any rational function model of the form of \cref{eq:model}, assuming the standard ranking described in \cref{ranking}. Generate $\tilde{M}$ by scaling any parameter $\tilde{\theta} \in \pmb{\theta}$ in the state variable equations of $M$ with $\tilde{u}$.}
     
\textit{Then for each input-output relation precursor equation $P'(M)$, $\tilde{M}$ will have an analogous input-output equation precursor $P'(\tilde{M})$, where the terms of $P'(M)$ will also be present in $P'(\tilde{M})$ with two modifications: 1) the replacement rule $\tilde{\theta} \mapsto \tilde{\theta}\tilde{u}$ (which may split a monomial into two or more distinct terms), and 2) the multiplication of all such analogous terms in $P'(\tilde{M})$ by a common term in the parameters and $\tilde{u}$. We describe these modifications in more detail below.}

\textit{Let $C_im_i = (c_0 + c_1\tilde{\theta} + \cdots c_n\tilde{\theta}^n)m_i$ be a monomial term in $P'(M)$, where $C_i$ is a coefficient polynomial in the parameters (potentially including $\tilde{\theta}$, written here as a polynomial in $\tilde{\theta}$ where some $c_i$ may be zero), and  $m_i$ is a monomial in outputs, inputs, and their derivatives. }

\textit{Then $P'(\tilde{M})$ will contain analogous terms to $C_im_i$, of the form $ac_0m_i$, $ac_1\tilde{\theta}\tilde{u}m_i$, $\dots$ $ac_n\tilde{\theta}^n\tilde{u}^nm_i$, where $a$ is a common multiplicative term given by a polynomial in the parameters multiplied by a monomial in $\tilde{u}$ ($a$ may be trivial, e.g. $a=1$).}

\textit{Additional terms not descended from any original term may also be generated in $P'(\tilde{M})$, but these will never be the leading term of $P'(\tilde{M})$.}

\vspace{0.2cm}

\begin{pf}
    Let $M$ be any model of the form of \cref{eq:model}. 
    We take the following ranking on the input, output, and state variables as described previously:
    \begin{multline}
        u_1 < u_2 < ... < \dot{u}_1 < \dot{u}_2 < ...\\  < y_1 < y_2< ... <\dot{y}_1 < \dot{y}_2 < ...\\  < x_1 < x_2 < ...<\dot{x}_1 < \dot{x}_2 < ...
    \end{multline}

    Assume $\tilde{u}$ is a known forcing input function.
    Scale any parameter $\tilde{\theta} \in \pmb{\theta}$ in the state equations of $M$ by $\tilde{u}$ to create a new model $\tilde{M}$.
    For our ranking of the input, output, and state variables for the new system, we will choose the following:
    \begin{multline}
        \mathcolorbox{Yellow}{\tilde{u}} < u_1 < u_2 < ...< \mathcolorbox{Yellow}{\dot{\tilde{u}}} < \dot{u}_1 < \dot{u}_2 < ... < ...\\  < y_1 < y_2 < ... <\dot{y}_1 < \dot{y}_2 < ... \\< x_1 < x_2 < ...<\dot{x}_1 < \dot{x}_2 < ...
    \end{multline}
    where the new forcing input function is ranked lowest among the input variables, and importantly, less than output variables and state variables. 
    (For this proof, we only need to guarantee that the new forcing input function $\tilde{u}$ is less than the output and state variables, as well as their derivatives, but this makes clear that $\tilde{u}$ will never be the leader of the equations in which it appears.)

    To prove our result, we will prove two points: first, that the introduction of $\tilde{u}$ will not change the order of operations taken by the characteristic set algorithm in terms of the order in which polynomial pairs are chosen to reduce. Second, we will show that at each step of the characteristic set algorithm, the monomial terms from $M$ will be present in the equivalent equations for $\tilde{M}$, with the modifications specified in the lemma statement.
    
    To do this, we will review the changes this introduction makes to each step of the characteristic set algorithm: Rewrite, Compare, Reduce and Replace (we need not worry about the Enforce Monic step since we are considering only the input-output precursor equation), using induction on the number of algorithm iterations: For the base case, the original polynomial structure is preserved since scaling only affects coefficients. For the inductive step, assume the property holds for the previous iteration. When taking the derivative using the Leibniz rule in the next step, we will demonstrate that the term where $\tilde{u}$ is not differentiated preserves the original structure. Additional terms involving derivatives of $\tilde{u}$ are generated but ranked lower, maintaining the reduction order.
    
    \vspace{0.2cm}
    \noindent \textit{Rewrite.} Since input functions rank below output and state variable functions, the introduction of a new forcing input function will not change the first step of Ritt's Pseudo-Division, ordering the polynomials by their leaders. 
    
    A leader is the highest ranking variable in a polynomial, which, by the ranking system and our equation structure, can never be an input.
    That is, where for $M$ the leader of a polynomial was $x_i$, for $\tilde{M}$ that equivalent polynomial may now have leader $x_i$ scaled by $\tilde{u}$, but this does not change the leader, or leading variable, of each equation.
    
    As we will see below, extra steps of the Reduce-Replace sub-loop may be required, but the polynomials compared for reduction during each iteration of Ritt's Pseudo-Division will not change between the reduction steps of $M$ and $\tilde{M}$.

    \vspace{0.2cm}
    \noindent \textit{Compare.} Suppose that the equations at the previous step follow the conditions described, i.e. each monomial in the equivalent step for the calculation for $M$ is present in $\tilde{M}$, but with the modification $\tilde{\theta} \mapsto \tilde{\theta}\tilde{u}$, and potentially multiplied by a coefficient in the parameters and $\tilde{u}$. Then because of the ranking chosen, note that the choice of $A_j$ and $A_i$ in the Compare step will proceed analogously, i.e. the same equations will be chosen here.
    
    During the Compare step, we may need to take the derivative of terms that now contain $\tilde{u}$. 
    Consider an arbitrary monomial of the polynomial $A_j$ given by $h_j(\pmb{\theta}) f_j(\pmb{x})$ where $h_j(\cdot)$ and $f_j(\cdot)$ are polynomials. 
    Assume the corresponding modified monomial in the augmented polynomial $\tilde{A}_j$ is represented by $h_j(\tilde{\theta} \tilde{u}; \pmb{\theta}\setminus \tilde{\theta}) a f_j(\pmb{x}) $, where now every instance of $\tilde{\theta}$ is scaled by a new function $\tilde{u}$, and $a$ is the multiplicative term described in the lemma statement which may include $\tilde{u}$.
    Then, taking the $k$-th derivative we find:
    \begin{fleqn}
    \begin{align}
        \big(h_j(\pmb{\theta}) f_j(\pmb{x})\big)^{(k)} =& h_j(\pmb{\theta}) f_j(\pmb{x})^{(k)} \label{orgder}
    \end{align}

    \begin{multline}
        \big(h_j(\tilde{\theta} \mathcolorbox{Yellow}{\tilde{u}}; \pmb{\theta}\setminus \tilde{\theta}) \mathcolorbox{Yellow}{a} f_j(\pmb{x})\big)^{(k)} 
        = \\ \sum_{m=0}^{k} {\binom{k}{m}} \big(h_j(\tilde{\theta} \mathcolorbox{Yellow}{\tilde{u}}; \pmb{\theta}\setminus \tilde{\theta}) \mathcolorbox{Yellow}{a} \big)^{(k-m)}f_j(\pmb{x})^{(m)}  
        \label{newder2}  
    \end{multline}
    \end{fleqn}
    where we have applied the Leibniz Rule and Chain Rule to find the $k$-th derivative for the modified term in \cref{newder2}. 

    The term corresponding to $m=k$ in \cref{newder2} is $h_j(\tilde{\theta} \mathcolorbox{Yellow}{\tilde{u}}; \pmb{\theta}\setminus \tilde{\theta}) \mathcolorbox{Yellow}{a} f_j(\pmb{x})^{(k)}$, which preserves the structure related to the $k$-th derivative of the term from the original system $M$, $h_j(\tilde{\theta}; \pmb{\theta}) f_j(\pmb{x})^{(k)}$ under the two modifications described above.
    Assuming $\tilde{u}$ is a constant function will recover the corresponding monomial term, $h_j(\pmb{\theta}) f_j(\pmb{x})^{(k)}$, of the original system $M$ up to multiplication by $a$.

    In addition to the original monomial $k$-th derivative term, extra monomial term(s) with potentially new information have been added to \cref{newder2}:
    \begin{itemize}
        \item Due to the ranking system and the fact that these new terms all contain derivatives of $\tilde{u}$, they will not combine with the previously existing analogous monomial terms from the characteristic set calculation of $M$ as they are distinct monomials. 
        \item During the Reduce step, these extra $k$ terms will be added to the pseudo-division remainder $R$, and therefore to $\tilde{A}_i$.
        \begin{itemize}
            \item If these new monomial terms contain derivatives of the leader $w_j$ of $\tilde{A}_j$, then during the current Reduce-Replace cycle, an extra $k$-sub-loops will be required to fully reduce $\tilde{A}_i$ with respect to $\tilde{A}_j$.
            \item If these new terms do \textit{not} contain derivatives of the current leader $w_j$, then during a later Reduce-Replace cycle where this is the case, an extra $k$-sub-loops will be required to fully reduce $\tilde{A}_i$ with respect to $\tilde{A}_j$.
        \end{itemize}
    \end{itemize}

    In this manner, the original terms and their derivatives from $M$ are represented in $\tilde{M}$ during the steps of Ritt's Pseudo-Division.
    Other terms involving derivatives of $\tilde{u}$ may be generated, but they will not alter the choice of leader for reduction steps based on $f_j(\pmb{x})$ and its derivatives, as these new terms will always be lower ranked.
    
    \vspace{0.2cm}
    \noindent \textit{Reduce and Replace.} 
    The multiply coefficient of the Reduce step may now contain $\tilde{u}$, if it is scaling the leader $w_j$ of the polynomial $\tilde{A}_j$, potentially leading to all terms being multiplied by some power of $\tilde{u}$ as described in the lemma statement.
    Note that even if we need to differentiate $\tilde{A}_j$ to find $w_j^{(k)}$, the coefficient of the leader $w_j$ does not change, as we can see from \cref{newder2}:
    the highest ranking derivative of the leader $w_j$ corresponds to the $m=k$ case, which always has $h_j(\tilde{\theta} \tilde{u}; \pmb{\theta}\setminus \tilde{\theta}) a$ as the coefficient.

    Recall from the definition of the Replace step that ``the sub-loop (Reduce and Replace) is iterated using $A_j^{(k-1)}$ in place of $A_j^{(k)}$ and so on, until the pseudo-remainder is reduced with respect to $A_j$''.
    Since the coefficient of the $w_j^{(k)}$ term is the same for all $k$, then the same coefficient is used as the multiplication coefficient during each sub-loop of Reduce-Replace.

    This will result in a scaling coefficient of the form $\big(h_j(\tilde{\theta} \tilde{u}; \pmb{\theta}\setminus \tilde{\theta})a\big)^{\ell+k}$ on each monomial term of the fully reduced $A_i$; where $\ell$ is the power of the scaling coefficient originally under model $M$, and $k$ is the number of additional sub-loops performed for the modified system $\tilde{M}$ to complete the same reduction step.

    The new scaling factor will apply to \textit{all} monomial terms of the modified $\tilde{A}_i$ polynomial, including any other state variables that will need to be reduced in later steps of Ritt's Pseudo-Division. 
    In this way, the new scaling factor will be carried through on all monomial terms to the final reduction step which forms the associated input-output relation polynomial(s).

    To summarize: any modified multiply coefficient containing $\tilde{u}$ may generate further sub-loops of Reduce-Replace. 
    This will scale all terms of all polynomials reduced against this polynomial during the characteristic set calculation by $\big(h_j(\tilde{\theta} \tilde{u}; \pmb{\theta}\setminus \tilde{\theta})a\big)^{\ell+k}$ instead of what was originally $h_j(\pmb{\theta})^{\ell}$ under model $M$.
    This does not change the order of the larger $n$-iterations of the characteristic set algorithm, as in which polynomials are compared for reduction; only the number of Reduce-Replace sub-loops required to completely reduce one polynomial with respect to another.

    By induction, we conclude that for all iterations of characteristic set calculation using Ritt's Pseudodivision algorithm, the fundamental polynomial reduction structure is preserved. Each iteration maintains the original parameter relationships under the two modifications given in the lemma statement, while potentially adding new lower-ranked terms involving the forcing function derivatives. $\square$
\end{pf}

\section*{Declaration of Competing Interest}
The authors declare that they have no known competing financial interests or personal relationships that could have appeared to influence the work reported in this paper.

\section*{Funding}
This work is the result of the research funded by the National Science Foundation  (grant no. DMS-1853032), the Clare Boothe Luce Program for Women in STEM, the Michigan Institute for Computational Discovery and Engineering, and Los Alamos National Laboratory Center for Nonlinear Studies.

\section*{Authorship contribution statement}
\textbf{Jessica Rose Conrad:} Conceptualization, Methodology, Software, Formal analysis, Investigation, Writing -- original draft, review \& editing.

\textbf{James M. Hyman:}  Methodology, Supervision, Writing -- review \& editing.

\textbf{Marisa C. Eisenberg:} Conceptualization, Methodology, Writing -- review \& editing, Supervision, Funding acquisition.

\section*{Data availability}
No data was used for the research described in this article. All mathematical examples and code used to generate figures are available either in the text or from the corresponding author upon reasonable request.

\section*{Use of AI tools}
The authors used AI assistance (Claude, Anthropic) for LaTeX formatting and document preparation. All mathematical results, proofs, theorems, biological interpretations, and scientific conclusions are entirely the work of the authors. AI was used solely for technical manuscript preparation and did not contribute to the research, analysis, or scientific content.

\end{document}